\documentclass[aps,prd,twocolumn,preprintnumbers,superscriptaddress,dblfloatfix,nofootinbib]{revtex4-1}
\usepackage{bbm}
\usepackage{mathrsfs}
\usepackage{slashed}
\usepackage{caption}
\usepackage{epstopdf}
\usepackage[normalem]{ulem}
\usepackage[bottom]{footmisc}
\usepackage{subcaption}
\usepackage{bbold}
\usepackage{threeparttable}
\usepackage{booktabs}
\usepackage{changepage}
\usepackage[utf8]{inputenc}

\usepackage{grffile}

\usepackage{graphicx}  
\usepackage{dcolumn}   
\usepackage{bm}        
\usepackage{amssymb}   
\usepackage{setspace}
\usepackage{amsmath, amssymb, setspace}
\usepackage{array}
\usepackage{booktabs}
\usepackage{caption}
\usepackage{indentfirst}
\usepackage{float}
\usepackage{lmodern}
\usepackage{multirow}
\usepackage{soul}
\usepackage[normalem]{ulem}

\usepackage{braket}
\usepackage{comment}

\usepackage{adjustbox} 
\newcommand{\DoBox}[1]{\begin{center}
\color{red}\fbox{
\begin{minipage}{0.9\textwidth}

\end{minipage}}
\end{center}}

\usepackage{color}

\begin{document}

\preprint{IPMU19-0090}

\title{Possible Hints of Sterile Neutrinos in Recent Measurements of the Hubble Parameter
}
 
\author{Graciela B. Gelmini}
\email{gelmini@physics.ucla.edu}
\affiliation{Department of Physics and Astronomy, University of California, 
Los Angeles, CA 90095-1547, USA}

\author{Alexander Kusenko}
 \email{kusenko@ucla.edu}
\affiliation{Department of Physics and Astronomy, University of California, 
Los Angeles, CA 90095-1547, USA}
\affiliation{Kavli Institute for the Physics and Mathematics of the Universe (WPI), UTIAS\\
The University of Tokyo, Kashiwa, Chiba 277-8583, Japan} 
 
\author{Volodymyr Takhistov}
\email{vtakhist@physics.ucla.edu}
\affiliation{Department of Physics and Astronomy, University of California, 
Los Angeles, CA 90095-1547, USA}
\affiliation{Kavli Institute for the Physics and Mathematics of the Universe (WPI), UTIAS\\
The University of Tokyo, Kashiwa, Chiba 277-8583, Japan}


\date{\today}

\begin{abstract}
Local Universe observations find a value of the Hubble constant $H_0$ that is larger than the value inferred from the Cosmic Microwave Background and other early Universe measurements, assuming
known physics and the $\Lambda$CDM cosmological model. We show that additional radiation in active neutrinos produced just before Big Bang Nucleosynthesis by an unstable sterile neutrino with mass $m_s=$ O(10) MeV can alleviate this discrepancy. The necessary masses and couplings of the sterile neutrino, assuming it mixes primarily with $\nu_{\tau}$ and/or $\nu_{\mu}$ neutrinos, are within reach of Super-Kamiokande as well as upcoming laboratory experiments such as NA62 and DUNE.
\end{abstract}

\maketitle

The standard $\Lambda$CDM cosmological model, in which the cosmological expansion is dominated by dark energy (a cosmological constant) and cold dark matter, has had multiple triumphs in explaining   cosmological observations. However, several discrepancies have persisted. The most notorious is the difference between the values of the Hubble constant $H_0$, which parametrizes the expansion rate of the Universe, as measured locally and those inferred from Cosmic Microwave Background (CMB) measurements.

 Cosmology can be strongly affected by the production and dynamics of particles beyond those of the Standard Model. Of particular interest are sterile neutrinos. Sterile or right-handed neutrinos are an essential part of the seesaw mechanism~\cite{Yanagida:1979as,GellMann:1980vs,Yanagida:1980xy,Minkowski:1977sc} that generates small masses for active neutrinos.  While the original seesaw mechanism invoked a large-scale mass, there are models in which a low mass scale is generated naturally~\cite{Kusenko:2010ik}. Sterile neutrinos provide an excellent probe of early Universe cosmology~\cite{Gelmini:2019esj,Gelmini:2019wfp,Gelmini:2019clw}.

A precise measurement of the Hubble constant provides essential information about the cosmological energy content and could further reveal significant new features related to fundamental physics~\cite{Suyu:2012ax}. Multiple independent  measurements of the Hubble constant in the local Universe (e.g. from Cepheids and Type-Ia supernovae), which are nearly independent of the cosmological expansion history, lead to a value $H_0$ = 74.03 $\pm$1.42 km s$^{-1}$ Mpc$^{-1}$~\cite{Riess:2019cxk}. In contrast, the value inferred from  Planck satellite CMB data and  Baryon Acoustic Oscillations (BAO) data, assuming that the standard $\Lambda$CDM cosmological model describes the expansion history since recombination,is $H_0$ = 67.66 $\pm$ 0.42 km s$^{- 1}$ Mpc$^{-1}$~\cite{Aghanim:2018eyx}. The discrepancy is significant at the $\sim 5 \sigma$ level~\cite{Wong:2019kwg}.
Underestimation of systematic effects does not seem responsible 
for this discrepancy~\cite{Follin:2017ljs,Dhawan:2017ywl,Shanks:2018rka,Riess:2018kzi,Bengaly:2018xko,vonMarttens:2018bvz,Arendse:2019hev}.
This substantiates a possible cosmological origin of the tension in the $H_0$ measurements. 

Proposed modifications of the standard cosmological model to explain the $H_0$ discrepancy alter either the late time Universe or the early pre-CMB emission cosmology.
Among the many proposals considered to diminish the $H_0$ tension are those based on vacuum phase transitions~\cite{DiValentino:2017rcr}, an early period of dark energy domination~\cite{Poulin:2018cxd}, a time-dependent dark energy with negative values~\cite{Dutta:2018vmq,Dutta:2019pio}, holographic dark energy~\cite{Guo:2018ans,Kumar:2019wfs}, quintessence~\cite{Miao:2018zpw,DiValentino:2019exe}, a charged dark matter sector~\cite{Ko:2017uyb}, acoustic dark matter~\cite{Raveri:2017jto}, cannibal or decaying dark matter~\cite{Buen-Abad:2018mas,Choi:2019jck}, axions~\cite{DEramo:2018vss}, interacting dark matter~\cite{Yang:2018ubt,Yang:2018pej,Yang:2018euj,Blinov:2019gcj}, decaying $Z^{\prime}$ gauge bosons \cite{Escudero:2019gzq},
 decaying dark matter~\cite{Anchordoqui:2015lqa,Buch:2016jjp,Bringmann:2018jpr,Pandey:2019plg,Doroshkevich:1985,Doroshkevich:1984ster,Vattis:2019efj}, new physics in dark energy sector~\cite{DiValentino:2019ffd,Pan:2019gop,Vagnozzi:2019ezj}, sterile neutrinos~\cite{Battye:2013xqa,Doroshkevich:1984ster}, Majorons~\cite{Arias-Aragon:2020qip} and models with primordial black holes~\cite{Flores:2020drq}.
 
A way to alleviate the Hubble tension that we pursue here is based on changing the amount of radiation consisting of active neutrinos during the time of production of the CMB. It has been found that an extra contribution of $\Delta N_{\rm eff}= + 0.4$ (see Ref.~\cite{Bernal:2016gxb} and Fig.~35 of Ref.~\cite{Aghanim:2018eyx})
to the effective number $N_{\rm eff}$ of relativistic neutrino species or other particles not coupled to photons, generically called ``dark radiation'', to the content of the Universe during the decoupling epoch alleviates the tension between these measurements.
In the standard cosmology only three active neutrino species contribute to this type of radiation and $ N_{\rm eff}= 3.046$~\cite{Mangano:2005cc}. At temperatures $T < 1$ MeV, after $e^+e^-$ annihilation, $N_{\rm eff}$ is defined as
\begin{equation} \label{eq:neffdef}
\rho_{\rm rad} = \Big[2 + \dfrac{7}{4}\Big(\dfrac{4}{11}\Big)^{4/3} N_{\rm eff}\Big] \dfrac{\pi^2}{30} T^4~,
\end{equation}
 and $\Delta N_{\rm eff} = N_{\rm eff} - 3.046$.

In this work we show that a sterile neutrino, with mass $m_s \simeq \mathcal{O}(30)$ MeV decaying just before Big Bang Nucleosynthesis (BBN)  can produce the desired increase in $N_{\rm eff}$ to alleviate the $H_0$ discrepancy\footnote{The observed light neutrino mass splittings and mixing angles impose a restrictive upper limit on the active-sterile neutrino mixing of sterile neutrinos in this mass range, unless cancellations (accidental or otherwise) happen among several sterile neutrinos. E.g. in the ``symmetry protected scenario”~\cite{Kersten:2007vk} (see e.g.~\cite{Drewes:2018gkc, Chrzaszcz:2019inj} and references therein)  there is no upper limit from light neutrino data on the active-sterile neutrino mixing of two almost degenerate massive sterile neutrinos with similar mixing (in magnitude). In this particular model the acceptable bands shown in Fig.~1 would move down in $\sin^2(\theta)$ by about a factor of 2~\cite{Sabti:2020yrt} which would not change our conclusions. }. This is different from the earlier proposals based on decaying dark matter~\cite{Anchordoqui:2015lqa,Buch:2016jjp,Bringmann:2018jpr,Pandey:2019plg,Doroshkevich:1985,Doroshkevich:1984ster,Vattis:2019efj} and sterile neutrinos~\cite{Battye:2013xqa,Doroshkevich:1984ster} in that the decaying particle is a heavy sterile neutrino, and the decay products that contribute to radiation are standard neutrinos, not some exotic particles.  Our scenario is compatible with all experimental limits  if the sterile neutrino couples primarily to either  $\nu_{\mu}$ or  $\nu_{\tau}$ active neutrinos. 

\begin{figure*}
\begin{center}
\includegraphics[width=0.32 \textwidth]{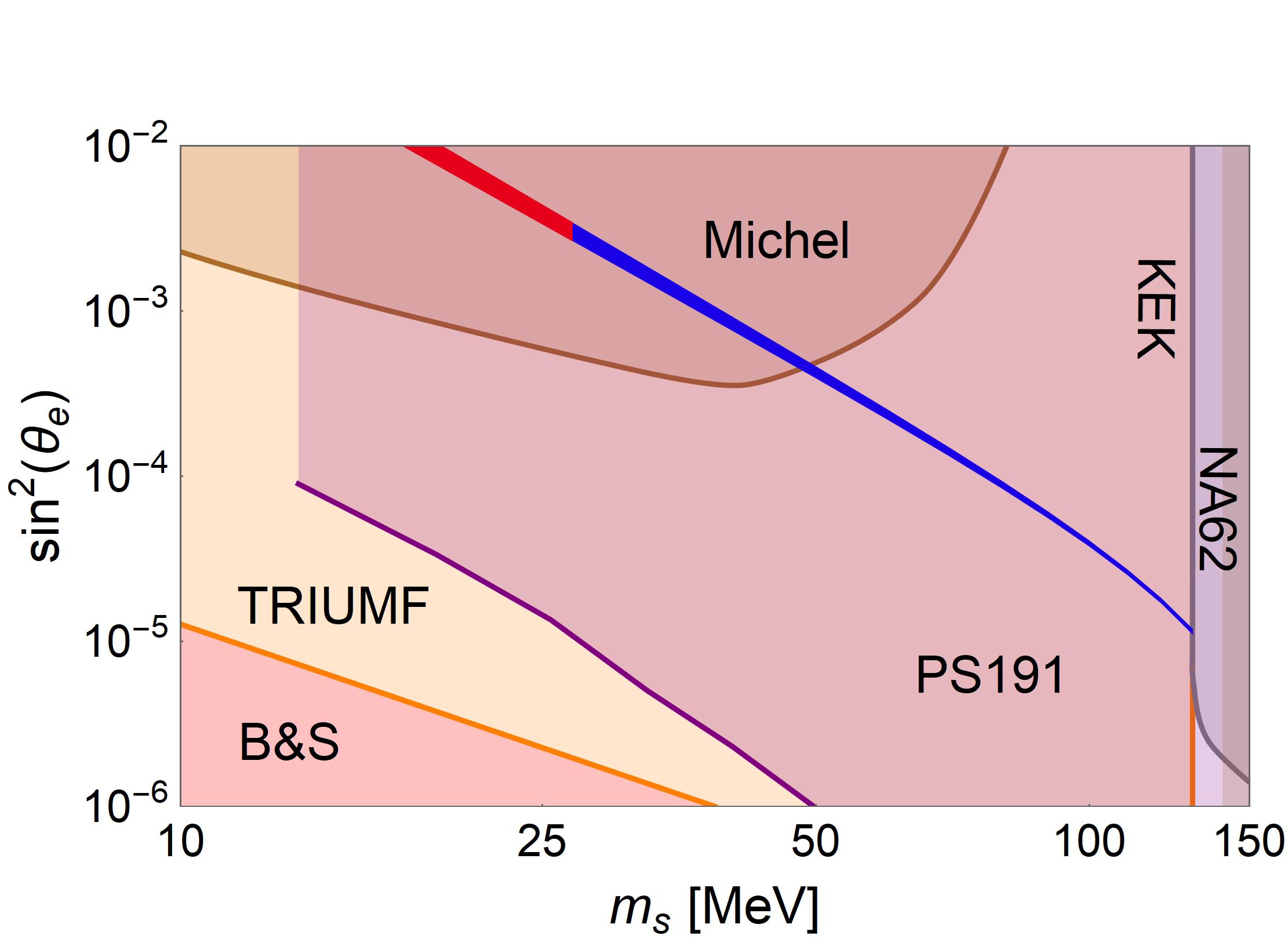}
\includegraphics[width=0.32 \textwidth]{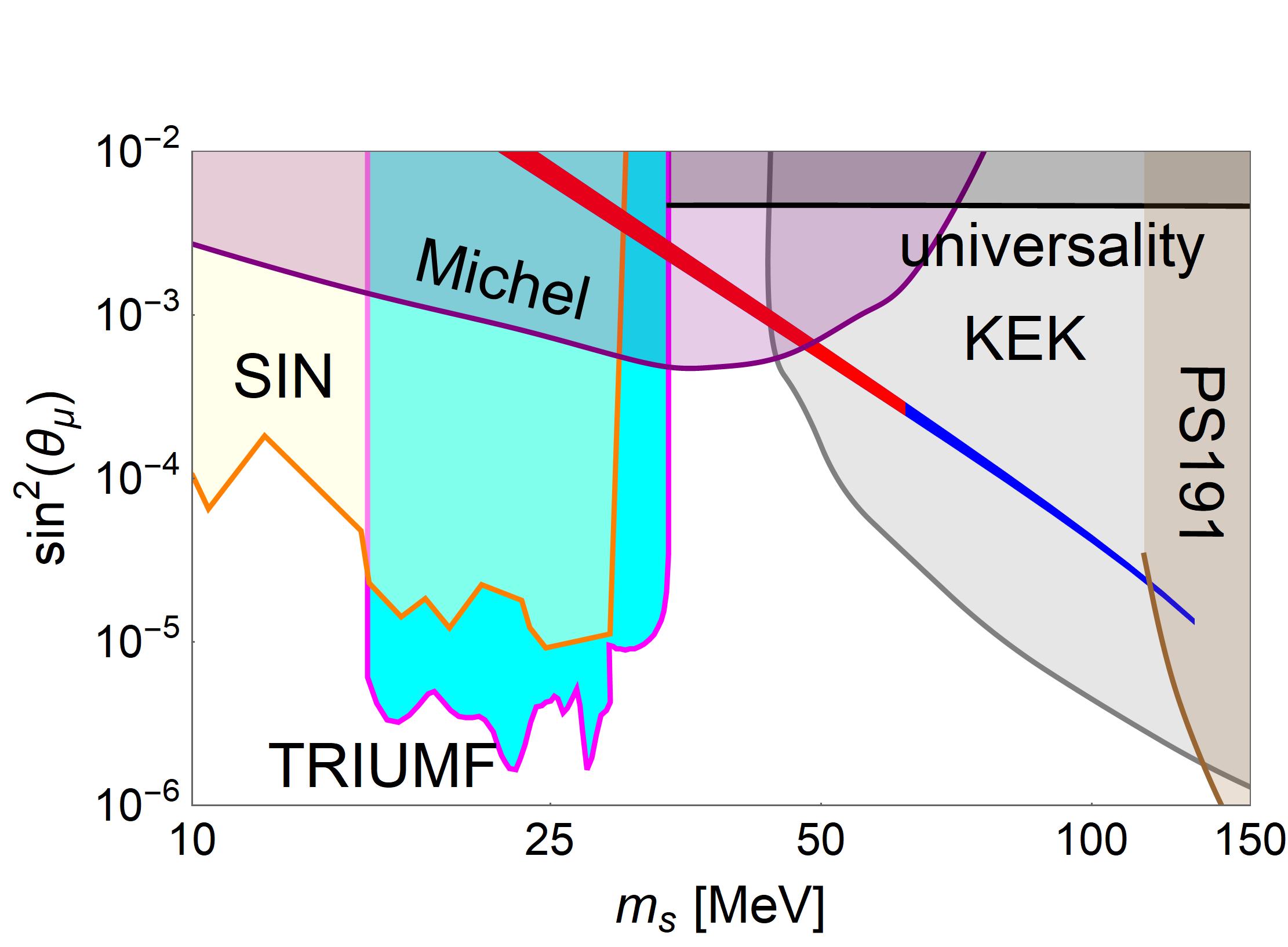}
\includegraphics[width=0.32  \textwidth]{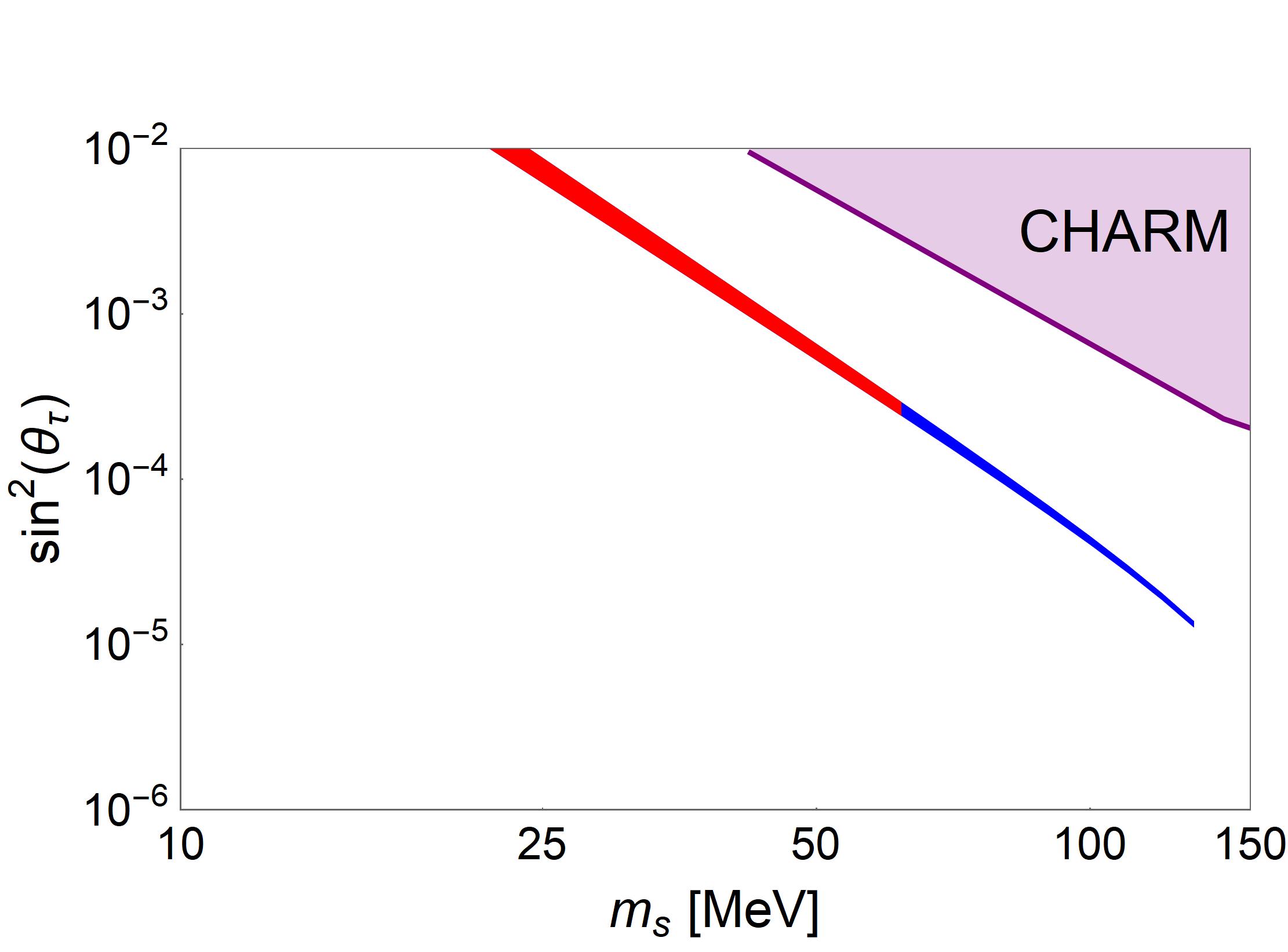}
\caption{Narrow (blue/red) bands in ($m_s, \sin^2\theta)$ space where decaying sterile neutrinos can alleviate the Hubble constant tension, assuming coupling exclusively to either electron neutrinos [left], or  muon neutrinos [center], or tau neutrinos [right]. Experimental limits from TRIUMF~\cite{PIENU:2011aa,Britton:1992,Britton:1992prl,Aguilar-Arevalo:2019owf}, KEK~\cite{Yamazaki:1984sj}, CHARM~\cite{Bergsma:1985is}, SIN~\cite{Abela:1981nf}, PS191~\cite{Bernardi:1985ny,Bernardi:1987ek}, 
Michel spectrum~\cite{deGouvea:2015euy}, 
lepton universality~\cite{deGouvea:2015euy}, NA62~\cite{NA62:2020mcv} and a re-analysis by Bryman and  Shrock~\cite{Bryman:2019ssi,Bryman:2019bjg},
Blue/red bands indicate regions where $\Delta N_{\rm eff} = 0.2-1$ is allowed according to Ref.~\cite{Dolgov:2000jw}. In the red portion of each band} Ref.~\cite{Ruchayskiy:2012si} finds that $\Delta N_{\rm eff} \leq 0.2$ is compatible with $Y_p < 0.264$.
\label{fig:lightheavyms}
\end{center}
\end{figure*} 

Let us consider a sterile neutrino $\nu_s$ with a mass $m_s$, which couples to the Standard Model particles only through its mixing $\sin \theta$ with an active neutrino $\nu_a$ ($a=e,\mu,\tau$).
Sterile neutrinos can be in equilibrium while they are relativistic  if, at some early time, their production rate $\Gamma_s$ is larger than the expansion rate of the Universe $H(T)$, resulting in the condition~\cite{Dolgov:2000jw}
\begin{equation}
    \dfrac{\Gamma_s}{H(T)} \simeq 2 \sin^2 \theta \Big(\dfrac{T}{3~\text{MeV}}\Big)^3 > 1~.
\end{equation}
Requiring that the temperature  $T > m_s$, one obtains 
\begin{equation}
    \sin^2 \theta > 5 \times 10^{-4} \Big(\dfrac{30~\text{MeV}}{m_s}\Big)^3~.
\end{equation}
 We will consider sterile neutrinos that satisfy this condition, and which are  in thermal equilibrium while they are  relativistic. They become non-relativistic and decay after the active neutrinos have decoupled, shortly before the beginning of BBN, at temperatures $T \gtrsim 1$ MeV and time $t \lesssim 1$ s. 

The effect of these sterile neutrinos on BBN has been studied in detail for sterile neutrino masses we are interested in, i.e. below the pion mass, i.e. $m_s < 135$ MeV (e.g. see Ref.~\cite{Dolgov:2000jw,Ruchayskiy:2012si,Sabti:2020yrt} and refs. therein). Recent re-examinations of this issue~\cite{Boyarsky:2020dzc, Sabti:2020yrt,Bondarenko:2021cpc} agree with
the results of Refs.~\cite{Dolgov:2000jw,Ruchayskiy:2012si} that we use in the following when considering one sterile neutrino.  Ref.~\cite{Sabti:2020yrt} considered also the case of two degenerate sterile neutrinos with similar active-sterile mixing and showed that the bands in Fig.~1 would be lower by about a factor of two in $\sin^2(\theta)$, which would not change our conclusions.  These sterile neutrinos decay primarily into active neutrinos  via the $\nu_s \rightarrow 3\nu_a$ channel, with a branching ratio of $2/3$. They also decay through $\nu_s \rightarrow \nu_a e^+e^-$ with a branching ratio of 1/3 as well as radiatively, via $\nu_s\rightarrow \nu_a\gamma$, with a suppressed branching ratio of $\mathcal{O}(10^{-2})$ (see e.g. Ref.~\cite{Fuller:2011qy,Atre:2009rg}). The corresponding lifetime $\tau_s$ of these sterile neutrinos is 
\begin{align}
\tau_s^{-1} \simeq&~ \Gamma_{3\nu} = \dfrac{G_F^2}{192 \pi^3} m_s^5 \sin^2 \theta  \notag\\
\simeq&~ 3.47 \times 10^{-5}~\text{s}^{-1} \Big(\dfrac{m_s}{\text{MeV}}\Big)^5 \sin^2 \theta~.
\end{align}

The heavy unstable neutrinos with $m_s \simeq \mathcal{O}(10)$ MeV that decay shortly prior to BBN lead to an increased production of $^4$He, because their decay products increase the energy density and thus speed up the expansion rate of the Universe. In turn, this results in an earlier freeze-out of the neutron to proton, $n/p$, ratio. In addition, electron neutrinos produced in the decay enter directly into the $p-n$ reactions and the small amount of $e^+e^-$ pairs produced change slightly the temperature of the photons, also affecting BBN.

In Fig.~\ref{fig:lightheavyms} we display the red/blue bands in the sterile neutrino mass and mixing  $(m_s, \sin^2 \theta)$ parameter space in which $\Delta N_{\rm eff}$ can be between 0.2 and 1 obtained in the analysis of Ref.~\cite{Dolgov:2000jw} (see Fig.~1 of Ref.~\cite{Dolgov:2000jw}). The bands are narrow and the value we are interested in, $\Delta N_{\rm eff}= 0.4$ is somewhere in the middle of it. We indicate in red the portions of the bands in which the more recent analysis of Ref.~\cite{Ruchayskiy:2012si} finds that $\Delta N_{\rm eff} \geq 0.2$ is compatible with the upper limit of $Y_p < 0.264$ on the primordial $^4$He abundance (see Fig.~3 of Ref.~\cite{Ruchayskiy:2012si}). The red/blue bands are shown overlapped with current limits assuming mixing of $\nu_s$ exclusively with either $\nu_e$, or $\nu_{\mu}$ or $\nu_{\tau}$, in the panels from left to right, respectively. We note that pion decay searches~\cite{PIENU:2011aa,Britton:1992,Britton:1992prl} already reject sterile neutrinos mixing primarily with $\nu_e$ in the desired region of the parameter space.
A recent re-analysis  of existing data by  Bryman and Shrock~\cite{Bryman:2019ssi} produced stronger upper limits on the sterile neutrino mixing with active electron neutrinos as well as muon neutrinos~\cite{Bryman:2019bjg}. An
even stronger upper limit can be derived on this mixing from
neutrino-less double-beta searches, if neutrinos are Majorana
particles~\cite{Benes:2005hn}, however, these stronger limits do not affect the
conclusion that the sterile neutrinos we propose should not
mix with electron neutrinos.
The case of sterile neutrinos mixing with muon neutrinos, which  is under pressure from a Michel spectrum analysis~\cite{deGouvea:2015euy} (and references therein), is within the expected reach of the Super-Kamiokande experiment~\cite{Kusenko:2004qc} as well as upcoming searches of other laboratory experiments such as NA62 \cite{Drewes:2018gkc} and the DUNE near detector~\cite{Ballett:2019bgd}. The sterile neutrino mixing with tau neutrinos is unconstrained in the band of interest  
and is also within reach of upcoming searches of laboratory experiments such as NA62 \cite{Drewes:2018gkc} and the DUNE near detector~\cite{Ballett:2019bgd}.
While additional limits derived from SN1987A data on sterile neutrinos have been suggested~(e.g. Refs.~\cite{Dolgov:2000jw,Dolgov:2000pj,Mastrototaro:2019vug}), they have been put into question by more comprehensive analyses (e.g. Refs.~\cite{Syvolap:2019dat,Suliga:2019bsq,Suliga:2020vpz}).

Each band of $\Delta N_{\rm eff}= 0.2-1.0$ in Fig.~\ref{fig:lightheavyms} (taken from Ref.~\cite{Dolgov:2000jw}) implies a band in the sterile neutrino lifetime as a function of $m_s$ (see Fig.~2 of Ref.~\cite{Dolgov:2000jw}) that falls within the lifetime band displayed in Fig.~2 of Ref.~\cite{Ruchayskiy:2012si} corresponding to the values of $Y_p$ in the range of 0.250-0.264~\cite{Izotov:2014fga} (and follows the $Y_p = 0.250$  line for $m_s < 60$ MeV). This shows that results of Ref.~\cite{Ruchayskiy:2012si} and Ref.~\cite{Dolgov:2000jw} are consistent\footnote{Eq.~(17) of Ref.~\cite{Ruchayskiy:2012si} gives a non-standard definition for $N_{\rm eff}$, without a factor of $(4/11)^{4/3}$ on the right hand side. However, the text of the paper clarifies that the usual definition of $N_{\rm eff}$, as given in our Eq.~\eqref{eq:neffdef} for temperatures $T < 1$ MeV,  is employed in the calculations.}.
The above mentioned range of $Y_p$, employed in Ref.~\cite{Ruchayskiy:2012si}, is similar to the $2\sigma$ range, $0.2507 < Y_p < 0.2595$, given more recently in Ref.~\cite{Izotov:2014fga}. This measurement is somewhat higher than other recent results, as discussed in the BBN review of the Particle Data Group (PDG)~\cite{Tanabashi:2018oca}.

An additional comment is in order regarding the primordial deuterium abundance, as employed in the analysis of Ref.~\cite{Ruchayskiy:2012si}. The $3\sigma$ range used in Ref.~\cite{Ruchayskiy:2012si}, $D/H = (2.2-3.5) \times 10^{-5}$, is more generous than the $3\sigma$ interval  adopted by PDG~\cite{Tanabashi:2018oca}, $D/H = (2.488-2.650)\times 10^{-5}$, but it is within the allowed range of values inferred from recent measurements (see e.g. Table 5 of  Ref.~\cite{Zavarygin:2017}). Comparing the two panels of the Fig.~3 of Ref.~\cite{Ruchayskiy:2012si} (showing the upper limits on  the deuterium abundance and on the $N_{\rm eff}$ imposed by $Y_p<0.264$, respectively) one can see that  $N_{\rm eff} \simeq 3.4$ corresponds to $D/H \simeq 2.8 \times 10^{-5}$, a less extreme value than the maximum, $3.5 \times 10^{-5}$, of the range the authors employ ($D/H \simeq 2.65 \times 10^{-5}$ corresponds to $N_{\rm eff} \simeq 3.2$ instead). In Ref.~\cite{Ruchayskiy:2012si} the authors note that the relation they find between $D/H$ and $N_{\rm eff}$ is the same as in a model without new particles and just with $N_{\rm eff}$ different from 3. In this regard, a maximum likelihood analysis based on current measurements of $^4$He and $D$ primordial abundances and allowing for just a change in $N_{\rm eff}$ and the baryon to photon ratio $\eta$ found that $2.3 < N_{\rm eff} < 3.4$ is allowed~\cite{Cyburt:2015mya,Tanabashi:2018oca}.

A simple estimate of the freeze-out temperature $T_{\rm f.o.}^R$ of sterile neutrinos assumed to be relativistic results from the condition $\Gamma(T_{\rm f.o.}^R) = H(T_{\rm f.o.}^R)$, where the interaction rate
$\Gamma= \sigma n \simeq G_F^2 T^5$
is equal to the Hubble expansion rate $H \simeq \sqrt{\rho_{\rm rad}}/{M_{\rm Pl}} \simeq T^2/M_{\rm Pl}$,
giving
\begin{equation}
T_{\rm f.o.}^R \simeq 15~\text{MeV} \Big( \dfrac{10^{-3}}{\sin^2 \theta}\Big)^{1/3}~.    
\end{equation}
Only sterile neutrinos $\nu_s$ that satisfy the condition $T_{\rm f.o.}^R > m_s$ would decouple while relativistic. However, this condition implies $\sin^2 \theta < 1.2 \times 10^{-4} (30~\text{MeV}/m_s)^3$, which is  smaller than the mixing angles in the bands of interest corresponding to each mass $m_s$. This implies that the sterile neutrinos capable of explaining the $H_0$ discrepancy decouple while they are non-relativistic, although their decoupling temperature $T_{\rm f.o.}^{NR}$  is not much smaller than their mass so that the Boltzmann suppression factor $\exp\{- m_s/T_{\rm f.o.}^{NR}\}$ in their abundance is not negligible but it is not very small. This suppression factor allows to explain why the sterile neutrinos with masses in the 10's of MeV, can decay after the active neutrinos decouple,  at $T_{\rm decay} < 3 $ MeV, i.e. with $m_s \gg T_{\rm decay}$, and still produce $\Delta N_{\rm eff} < 1$. This is important because most of the energy produced in the sterile neutrino decays is deposited into the active neutrino sector, except for the small ($\lesssim 10$\%) amount going into $e^+e^-$ and $\gamma$'s. Thus there is very little entropy dilution of the active neutrino contribution to the density of the Universe.

 In order to interact with the surrounding particles, the active neutrinos produced in sterile neutrino decays should have an energy $E_a$ larger than a re-thermalization energy value $E^{\rm rt}$. This value can be estimated using the cross-section $\sigma_a \simeq G_F^2 E_a T$. At the moment of decay, assumed to be instantaneous for simplicity, the decay rate equals the expansion rate, $H(T_{\rm decay}) = \Gamma_{\rm decay}$. The re-thermalization energy at decay, $E_d^{\rm rt}$, is such that
\begin{equation}
G_F^2 E_d^{\rm rt} T_{\rm decay}^4 \simeq H(T_{\rm decay})\simeq \Gamma_{\rm decay} ~.  
\end{equation}
Using the $\nu_s \rightarrow 3\nu_a$ channel decay rate (and $g_{\ast} = 10$ for the relativistic number of degrees of freedom)  we obtain
\begin{equation}
    E_d^{\rm rt} = \dfrac{0.55~\text{MeV}}{\sin^2 \theta} \Big(\dfrac{10~\text{MeV}}{m_s}\Big)^5~.
\end{equation}
The characteristic energy of the active neutrino decay products at the time of decay is a fraction of the parent sterile neutrino mass, $\sim m_s/3$. This initial energy is smaller than the energy necessary for re-thermalization at decay, i.e. $m_s/3 < E_d^{\rm rt}$, for
\begin{equation}
\sin^2 \theta < 1.6 \times 10^{-4} (30~\text{MeV}/m_s)^6~.
\end{equation}
Is is straightforward to demonstrate that, as the temperature decreases after the decay, for $T < T_{\rm decay}$, the thermalization energy increases as $T^{-1}$, $E^{\rm rt} = E_d^{\rm rt} (T_{\rm decay}/T)$.
On the other hand, the active neutrino decay product energy redshifts, $E_a = (m_s/3)(T/T_{\rm decay})$.
Thus, if the active neutrino does not thermalize at decay, $m_s/3 < E_d^{\rm rt}$, it will not thermalize at any later time, $E_a(T) < E^{\rm rt}(T)$. Since the condition on $\sin^2\theta$ derived above is satisfied by the sterile neutrinos in the parameter space of interest, practically all the energy in their decays goes into active neutrinos which are hotter than the standard relic active neutrinos.

The above considerations open a possible way to distinctly probe our scenario in future observations.
Relic neutrinos act as radiation while they are relativistic, but they turn into cold dark matter after the expansion of the universe makes them non-relativistic.  The co-moving free-streaming length of neutrinos increases while they are relativistic, then decreases, so that the co-moving wave number passes through a minimum $k_\nu$.  If neutrinos account for fraction $f_\nu=\Omega_\nu/\Omega_m$ of matter, the free streaming of neutrinos suppresses the small-scale matter power spectrum $P(k)$ on scales $k>k_\nu $ by a factor $\sim (1-f_\nu)^2$, which can be observable~\cite{Lesgourgues:2006nd,Font-Ribera:2013rwa,Bolliet:2019zuz}. In our case, an additional non-thermal  higher-energy component of neutrinos is contributed by the decay of the sterile neutrinos.  This increases the fraction $f_\nu$.  Furthermore, the additional population introduces a new scale  $\tilde{k}_\nu<k_\nu$ which corresponds to the (longer length) scale at which the additional neutrinos become non-relativistic.  The resulting redshift dependent suppression in P(k), if it could be observed, would provide an additional test of this scenario.

We briefly comment on sterile neutrinos heavier than the pion masses, $m_s \gtrsim 135$ MeV, which decay mostly into final states containing pions. 
Sterile neutrinos in the mass range $150~\text{MeV} < m_s < 450~\text{MeV}$ and $2 \times 10^{-13} < \sin^2 \theta < 2 \times 10^{-11}$ are subject to stringent bounds from BBN~\cite{Gelmini:2020ekg}. However, these bounds can be relaxed in particular models with a large lepton asymmetry, in which the decay of these heavier sterile neutrinos could also produce  $\Delta N_{\rm eff} = 0.1-0.4$ and thus ease the Hubble parameter tension~\cite{Gelmini:2020ekg}.

Most of our results are based on Refs.~\cite{Dolgov:2000jw, Ruchayskiy:2012si}.  A recent work of Ref.~\cite{Boyarsky:2021yoh} calls these results into question,  implying that our scenario works only for tau neutrino mixing. However, non-equilibrium BBN dynamics are complicated; a critique and comparison of the two approaches can be found in Ref.~\cite{Mastrototaro:2021wzl}. 

In summary, we have shown that sterile neutrinos coupling to the Standard Model particles only through their mixing with active neutrinos, and decaying into regular Standard Model particles just before BBN, can alleviate the tension arising between the local and the early Universe measurements of the Hubble constant. The decays of the sterile neutrinos create a dark radiation in the form of active neutrinos which are hotter than the standard neutrino background.  This results in an increase in the effective number of relativistic neutrino species during the CMB emission epoch. Since the decays occur prior to BBN, they do not affect the CMB other than by changing $N_{\rm eff}$. 

 In particular, we find that sterile neutrinos with masses $m_s= O(30)$ MeV and mixing angles $\sin^2 \theta > 10^{-5}$ in the red/blue bands shown in Fig.~1,  that couple primarily to  $\nu_{\tau}$, or possibly $\nu_{\mu}$, active neutrinos, are compatible with BBN limits and can increase $N_{\rm eff}$ by $\Delta N_{\rm eff} \simeq 0.4$. This very minimal proposal, with only sterile neutrinos added to the SM, can be probed in existing terrestrial laboratories, such as Super-Kamiokande, as well as upcoming data from NA62, DUNE and other experiments. 
Furthermore, upcoming measurements of CMB-S4~\cite{Abazajian:2016yjj}, which are expected to probe $\Delta N_{\rm eff}$ at the few-percent level, will also test this scenario.

\section*{Acknowledgements}

This work was supported in part by the U.S. Department of Energy (DOE) Grant No. DE-SC0009937.  A.K. and V.T. also acknowledge support by the World Premier International Research Center Initiative (WPI), MEXT, Japan. A.K. and V.T. thank Aspen Center for Physics, which is supported by National Science Foundation grant PHY-1607611.

\bibliography{sterileH0bib}

\begin{thebibliography}{94}%
\makeatletter
\providecommand \@ifxundefined [1]{%
 \@ifx{#1\undefined}
}%
\providecommand \@ifnum [1]{%
 \ifnum #1\expandafter \@firstoftwo
 \else \expandafter \@secondoftwo
 \fi
}%
\providecommand \@ifx [1]{%
 \ifx #1\expandafter \@firstoftwo
 \else \expandafter \@secondoftwo
 \fi
}%
\providecommand \natexlab [1]{#1}%
\providecommand \enquote  [1]{``#1''}%
\providecommand \bibnamefont  [1]{#1}%
\providecommand \bibfnamefont [1]{#1}%
\providecommand \citenamefont [1]{#1}%
\providecommand \href@noop [0]{\@secondoftwo}%
\providecommand \href [0]{\begingroup \@sanitize@url \@href}%
\providecommand \@href[1]{\@@startlink{#1}\@@href}%
\providecommand \@@href[1]{\endgroup#1\@@endlink}%
\providecommand \@sanitize@url [0]{\catcode `\\12\catcode `\$12\catcode
  `\&12\catcode `\#12\catcode `\^12\catcode `\_12\catcode `\%12\relax}%
\providecommand \@@startlink[1]{}%
\providecommand \@@endlink[0]{}%
\providecommand \url  [0]{\begingroup\@sanitize@url \@url }%
\providecommand \@url [1]{\endgroup\@href {#1}{\urlprefix }}%
\providecommand \urlprefix  [0]{URL }%
\providecommand \Eprint [0]{\href }%
\providecommand \doibase [0]{http://dx.doi.org/}%
\providecommand \selectlanguage [0]{\@gobble}%
\providecommand \bibinfo  [0]{\@secondoftwo}%
\providecommand \bibfield  [0]{\@secondoftwo}%
\providecommand \translation [1]{[#1]}%
\providecommand \BibitemOpen [0]{}%
\providecommand \bibitemStop [0]{}%
\providecommand \bibitemNoStop [0]{.\EOS\space}%
\providecommand \EOS [0]{\spacefactor3000\relax}%
\providecommand \BibitemShut  [1]{\csname bibitem#1\endcsname}%
\let\auto@bib@innerbib\@empty
\bibitem [{\citenamefont {Yanagida}(1979)}]{Yanagida:1979as}%
  \BibitemOpen
  \bibfield  {author} {\bibinfo {author} {\bibfnamefont {T.}~\bibnamefont
  {Yanagida}},\ }\href@noop {} {\bibfield  {journal} {\bibinfo  {journal}
  {Proceedings of the Workshop on the Baryon Number of the Universe and Unified
  Theories, Tsukuba, Japan, 13-14 Feb 1979}\ } (\bibinfo {year} {1979})},\
  \bibinfo {note} {in Proceedings of the Workshop on the Baryon Number of the
  Universe and Unified Theories, Tsukuba, Japan, 13-14 Feb 1979}\BibitemShut
  {NoStop}%
\bibitem [{\citenamefont {Gell-Mann}\ \emph {et~al.}(1980)\citenamefont
  {Gell-Mann}, \citenamefont {Ramond},\ and\ \citenamefont
  {Slansky}}]{GellMann:1980vs}%
  \BibitemOpen
  \bibfield  {author} {\bibinfo {author} {\bibfnamefont {M.}~\bibnamefont
  {Gell-Mann}}, \bibinfo {author} {\bibfnamefont {P.}~\bibnamefont {Ramond}}, \
  and\ \bibinfo {author} {\bibfnamefont {R.}~\bibnamefont {Slansky}},\
  }\href@noop {} {\  (\bibinfo {year} {1980})},\ \bibinfo {note} {print-80-0576
  (CERN)}\BibitemShut {NoStop}%
\bibitem [{\citenamefont {Yanagida}(1980)}]{Yanagida:1980xy}%
  \BibitemOpen
  \bibfield  {author} {\bibinfo {author} {\bibfnamefont {T.}~\bibnamefont
  {Yanagida}},\ }\href@noop {} {\bibfield  {journal} {\bibinfo  {journal}
  {Prog. Theor. Phys.}\ }\textbf {\bibinfo {volume} {64}},\ \bibinfo {pages}
  {1103} (\bibinfo {year} {1980})}\BibitemShut {NoStop}%
\bibitem [{\citenamefont {Minkowski}(1977)}]{Minkowski:1977sc}%
  \BibitemOpen
  \bibfield  {author} {\bibinfo {author} {\bibfnamefont {P.}~\bibnamefont
  {Minkowski}},\ }\href@noop {} {\bibfield  {journal} {\bibinfo  {journal}
  {Phys. Lett.}\ }\textbf {\bibinfo {volume} {B67}},\ \bibinfo {pages} {421}
  (\bibinfo {year} {1977})}\BibitemShut {NoStop}%
\bibitem [{\citenamefont {Kusenko}\ \emph {et~al.}(2010)\citenamefont
  {Kusenko}, \citenamefont {Takahashi},\ and\ \citenamefont
  {Yanagida}}]{Kusenko:2010ik}%
  \BibitemOpen
  \bibfield  {author} {\bibinfo {author} {\bibfnamefont {A.}~\bibnamefont
  {Kusenko}}, \bibinfo {author} {\bibfnamefont {F.}~\bibnamefont {Takahashi}},
  \ and\ \bibinfo {author} {\bibfnamefont {T.~T.}\ \bibnamefont {Yanagida}},\
  }\href {\doibase 10.1016/j.physletb.2010.08.031} {\bibfield  {journal}
  {\bibinfo  {journal} {Phys. Lett.}\ }\textbf {\bibinfo {volume} {B693}},\
  \bibinfo {pages} {144} (\bibinfo {year} {2010})},\ \Eprint
  {http://arxiv.org/abs/1006.1731} {arXiv:1006.1731 [hep-ph]} \BibitemShut
  {NoStop}%
\bibitem [{\citenamefont {Gelmini}\ \emph
  {et~al.}(2020{\natexlab{a}})\citenamefont {Gelmini}, \citenamefont {Lu},\
  and\ \citenamefont {Takhistov}}]{Gelmini:2019esj}%
  \BibitemOpen
  \bibfield  {author} {\bibinfo {author} {\bibfnamefont {G.~B.}\ \bibnamefont
  {Gelmini}}, \bibinfo {author} {\bibfnamefont {P.}~\bibnamefont {Lu}}, \ and\
  \bibinfo {author} {\bibfnamefont {V.}~\bibnamefont {Takhistov}},\ }\href
  {\doibase 10.1016/j.physletb.2019.135113} {\bibfield  {journal} {\bibinfo
  {journal} {Phys. Lett. B}\ }\textbf {\bibinfo {volume} {800}},\ \bibinfo
  {pages} {135113} (\bibinfo {year} {2020}{\natexlab{a}})},\ \Eprint
  {http://arxiv.org/abs/1909.04168} {arXiv:1909.04168 [hep-ph]} \BibitemShut
  {NoStop}%
\bibitem [{\citenamefont {Gelmini}\ \emph {et~al.}(2019)\citenamefont
  {Gelmini}, \citenamefont {Lu},\ and\ \citenamefont
  {Takhistov}}]{Gelmini:2019wfp}%
  \BibitemOpen
  \bibfield  {author} {\bibinfo {author} {\bibfnamefont {G.~B.}\ \bibnamefont
  {Gelmini}}, \bibinfo {author} {\bibfnamefont {P.}~\bibnamefont {Lu}}, \ and\
  \bibinfo {author} {\bibfnamefont {V.}~\bibnamefont {Takhistov}},\ }\href
  {\doibase 10.1088/1475-7516/2019/12/047} {\bibfield  {journal} {\bibinfo
  {journal} {JCAP}\ }\textbf {\bibinfo {volume} {12}},\ \bibinfo {pages} {047}
  (\bibinfo {year} {2019})},\ \Eprint {http://arxiv.org/abs/1909.13328}
  {arXiv:1909.13328 [hep-ph]} \BibitemShut {NoStop}%
\bibitem [{\citenamefont {Gelmini}\ \emph
  {et~al.}(2020{\natexlab{b}})\citenamefont {Gelmini}, \citenamefont {Lu},\
  and\ \citenamefont {Takhistov}}]{Gelmini:2019clw}%
  \BibitemOpen
  \bibfield  {author} {\bibinfo {author} {\bibfnamefont {G.~B.}\ \bibnamefont
  {Gelmini}}, \bibinfo {author} {\bibfnamefont {P.}~\bibnamefont {Lu}}, \ and\
  \bibinfo {author} {\bibfnamefont {V.}~\bibnamefont {Takhistov}},\ }\href
  {\doibase 10.1088/1475-7516/2020/06/008} {\bibfield  {journal} {\bibinfo
  {journal} {JCAP}\ }\textbf {\bibinfo {volume} {06}},\ \bibinfo {pages} {008}
  (\bibinfo {year} {2020}{\natexlab{b}})},\ \Eprint
  {http://arxiv.org/abs/1911.03398} {arXiv:1911.03398 [hep-ph]} \BibitemShut
  {NoStop}%
\bibitem [{\citenamefont {Suyu}\ \emph {et~al.}(2012)\citenamefont {Suyu} \emph
  {et~al.}}]{Suyu:2012ax}%
  \BibitemOpen
  \bibfield  {author} {\bibinfo {author} {\bibfnamefont {S.~H.}\ \bibnamefont
  {Suyu}} \emph {et~al.},\ }\href@noop {} {\  (\bibinfo {year} {2012})},\
  \Eprint {http://arxiv.org/abs/1202.4459} {arXiv:1202.4459 [astro-ph.CO]}
  \BibitemShut {NoStop}%
\bibitem [{\citenamefont {Riess}\ \emph {et~al.}(2019)\citenamefont {Riess},
  \citenamefont {Casertano}, \citenamefont {Yuan}, \citenamefont {Macri},\ and\
  \citenamefont {Scolnic}}]{Riess:2019cxk}%
  \BibitemOpen
  \bibfield  {author} {\bibinfo {author} {\bibfnamefont {A.~G.}\ \bibnamefont
  {Riess}}, \bibinfo {author} {\bibfnamefont {S.}~\bibnamefont {Casertano}},
  \bibinfo {author} {\bibfnamefont {W.}~\bibnamefont {Yuan}}, \bibinfo {author}
  {\bibfnamefont {L.~M.}\ \bibnamefont {Macri}}, \ and\ \bibinfo {author}
  {\bibfnamefont {D.}~\bibnamefont {Scolnic}},\ }\href {\doibase
  10.3847/1538-4357/ab1422} {\bibfield  {journal} {\bibinfo  {journal}
  {Astrophys. J.}\ }\textbf {\bibinfo {volume} {876}},\ \bibinfo {pages} {85}
  (\bibinfo {year} {2019})},\ \Eprint {http://arxiv.org/abs/1903.07603}
  {arXiv:1903.07603 [astro-ph.CO]} \BibitemShut {NoStop}%
\bibitem [{\citenamefont {Aghanim}\ \emph {et~al.}(2018)\citenamefont {Aghanim}
  \emph {et~al.}}]{Aghanim:2018eyx}%
  \BibitemOpen
  \bibfield  {author} {\bibinfo {author} {\bibfnamefont {N.}~\bibnamefont
  {Aghanim}} \emph {et~al.} (\bibinfo {collaboration} {Planck}),\ }\href@noop
  {} {\  (\bibinfo {year} {2018})},\ \Eprint {http://arxiv.org/abs/1807.06209}
  {arXiv:1807.06209 [astro-ph.CO]} \BibitemShut {NoStop}%
\bibitem [{\citenamefont {Wong}\ \emph {et~al.}(2019)\citenamefont {Wong} \emph
  {et~al.}}]{Wong:2019kwg}%
  \BibitemOpen
  \bibfield  {author} {\bibinfo {author} {\bibfnamefont {K.~C.}\ \bibnamefont
  {Wong}} \emph {et~al.},\ }\href {\doibase 10.1093/mnras/stz3094} {\
  (\bibinfo {year} {2019}),\ 10.1093/mnras/stz3094},\ \Eprint
  {http://arxiv.org/abs/1907.04869} {arXiv:1907.04869 [astro-ph.CO]}
  \BibitemShut {NoStop}%
\bibitem [{\citenamefont {Follin}\ and\ \citenamefont
  {Knox}(2018)}]{Follin:2017ljs}%
  \BibitemOpen
  \bibfield  {author} {\bibinfo {author} {\bibfnamefont {B.}~\bibnamefont
  {Follin}}\ and\ \bibinfo {author} {\bibfnamefont {L.}~\bibnamefont {Knox}},\
  }\href {\doibase 10.1093/mnras/sty720} {\bibfield  {journal} {\bibinfo
  {journal} {Mon. Not. Roy. Astron. Soc.}\ }\textbf {\bibinfo {volume} {477}},\
  \bibinfo {pages} {4534} (\bibinfo {year} {2018})},\ \Eprint
  {http://arxiv.org/abs/1707.01175} {arXiv:1707.01175 [astro-ph.CO]}
  \BibitemShut {NoStop}%
\bibitem [{\citenamefont {Dhawan}\ \emph {et~al.}(2018)\citenamefont {Dhawan},
  \citenamefont {Jha},\ and\ \citenamefont {Leibundgut}}]{Dhawan:2017ywl}%
  \BibitemOpen
  \bibfield  {author} {\bibinfo {author} {\bibfnamefont {S.}~\bibnamefont
  {Dhawan}}, \bibinfo {author} {\bibfnamefont {S.~W.}\ \bibnamefont {Jha}}, \
  and\ \bibinfo {author} {\bibfnamefont {B.}~\bibnamefont {Leibundgut}},\
  }\href {\doibase 10.1051/0004-6361/201731501} {\bibfield  {journal} {\bibinfo
   {journal} {Astron. Astrophys.}\ }\textbf {\bibinfo {volume} {609}},\
  \bibinfo {pages} {A72} (\bibinfo {year} {2018})},\ \Eprint
  {http://arxiv.org/abs/1707.00715} {arXiv:1707.00715 [astro-ph.CO]}
  \BibitemShut {NoStop}%
\bibitem [{\citenamefont {Shanks}\ \emph {et~al.}(2019)\citenamefont {Shanks},
  \citenamefont {Hogarth},\ and\ \citenamefont {Metcalfe}}]{Shanks:2018rka}%
  \BibitemOpen
  \bibfield  {author} {\bibinfo {author} {\bibfnamefont {T.}~\bibnamefont
  {Shanks}}, \bibinfo {author} {\bibfnamefont {L.}~\bibnamefont {Hogarth}}, \
  and\ \bibinfo {author} {\bibfnamefont {N.}~\bibnamefont {Metcalfe}},\ }\href
  {\doibase 10.1093/mnrasl/sly239} {\bibfield  {journal} {\bibinfo  {journal}
  {Mon. Not. Roy. Astron. Soc.}\ }\textbf {\bibinfo {volume} {484}},\ \bibinfo
  {pages} {L64} (\bibinfo {year} {2019})},\ \Eprint
  {http://arxiv.org/abs/1810.02595} {arXiv:1810.02595 [astro-ph.CO]}
  \BibitemShut {NoStop}%
\bibitem [{\citenamefont {Riess}\ \emph {et~al.}(2018)\citenamefont {Riess},
  \citenamefont {Casertano}, \citenamefont {Kenworthy}, \citenamefont
  {Scolnic},\ and\ \citenamefont {Macri}}]{Riess:2018kzi}%
  \BibitemOpen
  \bibfield  {author} {\bibinfo {author} {\bibfnamefont {A.~G.}\ \bibnamefont
  {Riess}}, \bibinfo {author} {\bibfnamefont {S.}~\bibnamefont {Casertano}},
  \bibinfo {author} {\bibfnamefont {D.}~\bibnamefont {Kenworthy}}, \bibinfo
  {author} {\bibfnamefont {D.}~\bibnamefont {Scolnic}}, \ and\ \bibinfo
  {author} {\bibfnamefont {L.}~\bibnamefont {Macri}},\ }\href@noop {} {\
  (\bibinfo {year} {2018})},\ \Eprint {http://arxiv.org/abs/1810.03526}
  {arXiv:1810.03526 [astro-ph.CO]} \BibitemShut {NoStop}%
\bibitem [{\citenamefont {Bengaly}\ \emph {et~al.}(2018)\citenamefont
  {Bengaly}, \citenamefont {Andrade},\ and\ \citenamefont
  {Alcaniz}}]{Bengaly:2018xko}%
  \BibitemOpen
  \bibfield  {author} {\bibinfo {author} {\bibfnamefont {C.~A.~P.}\
  \bibnamefont {Bengaly}}, \bibinfo {author} {\bibfnamefont {U.}~\bibnamefont
  {Andrade}}, \ and\ \bibinfo {author} {\bibfnamefont {J.~S.}\ \bibnamefont
  {Alcaniz}},\ }\href@noop {} {\  (\bibinfo {year} {2018})},\ \Eprint
  {http://arxiv.org/abs/1810.04966} {arXiv:1810.04966 [astro-ph.CO]}
  \BibitemShut {NoStop}%
\bibitem [{\citenamefont {von Marttens}\ \emph {et~al.}(2019)\citenamefont {von
  Marttens}, \citenamefont {Marra}, \citenamefont {Casarini}, \citenamefont
  {Gonzalez},\ and\ \citenamefont {Alcaniz}}]{vonMarttens:2018bvz}%
  \BibitemOpen
  \bibfield  {author} {\bibinfo {author} {\bibfnamefont {R.}~\bibnamefont {von
  Marttens}}, \bibinfo {author} {\bibfnamefont {V.}~\bibnamefont {Marra}},
  \bibinfo {author} {\bibfnamefont {L.}~\bibnamefont {Casarini}}, \bibinfo
  {author} {\bibfnamefont {J.~E.}\ \bibnamefont {Gonzalez}}, \ and\ \bibinfo
  {author} {\bibfnamefont {J.}~\bibnamefont {Alcaniz}},\ }\href {\doibase
  10.1103/PhysRevD.99.043521} {\bibfield  {journal} {\bibinfo  {journal} {Phys.
  Rev.}\ }\textbf {\bibinfo {volume} {D99}},\ \bibinfo {pages} {043521}
  (\bibinfo {year} {2019})},\ \Eprint {http://arxiv.org/abs/1812.02333}
  {arXiv:1812.02333 [astro-ph.CO]} \BibitemShut {NoStop}%
\bibitem [{\citenamefont {Arendse}\ \emph {et~al.}(2020)\citenamefont {Arendse}
  \emph {et~al.}}]{Arendse:2019hev}%
  \BibitemOpen
  \bibfield  {author} {\bibinfo {author} {\bibfnamefont {N.}~\bibnamefont
  {Arendse}} \emph {et~al.},\ }\href {\doibase 10.1051/0004-6361/201936720}
  {\bibfield  {journal} {\bibinfo  {journal} {Astron. Astrophys.}\ }\textbf
  {\bibinfo {volume} {639}},\ \bibinfo {pages} {A57} (\bibinfo {year}
  {2020})},\ \Eprint {http://arxiv.org/abs/1909.07986} {arXiv:1909.07986
  [astro-ph.CO]} \BibitemShut {NoStop}%
\bibitem [{\citenamefont {Di~Valentino}\ \emph {et~al.}(2018)\citenamefont
  {Di~Valentino}, \citenamefont {Linder},\ and\ \citenamefont
  {Melchiorri}}]{DiValentino:2017rcr}%
  \BibitemOpen
  \bibfield  {author} {\bibinfo {author} {\bibfnamefont {E.}~\bibnamefont
  {Di~Valentino}}, \bibinfo {author} {\bibfnamefont {E.~V.}\ \bibnamefont
  {Linder}}, \ and\ \bibinfo {author} {\bibfnamefont {A.}~\bibnamefont
  {Melchiorri}},\ }\href {\doibase 10.1103/PhysRevD.97.043528} {\bibfield
  {journal} {\bibinfo  {journal} {Phys. Rev.}\ }\textbf {\bibinfo {volume}
  {D97}},\ \bibinfo {pages} {043528} (\bibinfo {year} {2018})},\ \Eprint
  {http://arxiv.org/abs/1710.02153} {arXiv:1710.02153 [astro-ph.CO]}
  \BibitemShut {NoStop}%
\bibitem [{\citenamefont {Poulin}\ \emph {et~al.}(2018)\citenamefont {Poulin},
  \citenamefont {Smith}, \citenamefont {Karwal},\ and\ \citenamefont
  {Kamionkowski}}]{Poulin:2018cxd}%
  \BibitemOpen
  \bibfield  {author} {\bibinfo {author} {\bibfnamefont {V.}~\bibnamefont
  {Poulin}}, \bibinfo {author} {\bibfnamefont {T.~L.}\ \bibnamefont {Smith}},
  \bibinfo {author} {\bibfnamefont {T.}~\bibnamefont {Karwal}}, \ and\ \bibinfo
  {author} {\bibfnamefont {M.}~\bibnamefont {Kamionkowski}},\ }\href@noop {} {\
   (\bibinfo {year} {2018})},\ \Eprint {http://arxiv.org/abs/1811.04083}
  {arXiv:1811.04083 [astro-ph.CO]} \BibitemShut {NoStop}%
\bibitem [{\citenamefont {Dutta}\ \emph {et~al.}(2018)\citenamefont {Dutta},
  \citenamefont {Ruchika}, \citenamefont {Roy}, \citenamefont {Sen},\ and\
  \citenamefont {Sheikh-Jabbari}}]{Dutta:2018vmq}%
  \BibitemOpen
  \bibfield  {author} {\bibinfo {author} {\bibfnamefont {K.}~\bibnamefont
  {Dutta}}, \bibinfo {author} {\bibnamefont {Ruchika}}, \bibinfo {author}
  {\bibfnamefont {A.}~\bibnamefont {Roy}}, \bibinfo {author} {\bibfnamefont
  {A.~A.}\ \bibnamefont {Sen}}, \ and\ \bibinfo {author} {\bibfnamefont
  {M.~M.}\ \bibnamefont {Sheikh-Jabbari}},\ }\href@noop {} {\  (\bibinfo {year}
  {2018})},\ \Eprint {http://arxiv.org/abs/1808.06623} {arXiv:1808.06623
  [astro-ph.CO]} \BibitemShut {NoStop}%
\bibitem [{\citenamefont {Dutta}\ \emph {et~al.}(2019)\citenamefont {Dutta},
  \citenamefont {Roy}, \citenamefont {Ruchika}, \citenamefont {Sen},\ and\
  \citenamefont {Sheikh-Jabbari}}]{Dutta:2019pio}%
  \BibitemOpen
  \bibfield  {author} {\bibinfo {author} {\bibfnamefont {K.}~\bibnamefont
  {Dutta}}, \bibinfo {author} {\bibfnamefont {A.}~\bibnamefont {Roy}}, \bibinfo
  {author} {\bibnamefont {Ruchika}}, \bibinfo {author} {\bibfnamefont {A.~A.}\
  \bibnamefont {Sen}}, \ and\ \bibinfo {author} {\bibfnamefont {M.~M.}\
  \bibnamefont {Sheikh-Jabbari}},\ }\href@noop {} {\  (\bibinfo {year}
  {2019})},\ \Eprint {http://arxiv.org/abs/1908.07267} {arXiv:1908.07267
  [astro-ph.CO]} \BibitemShut {NoStop}%
\bibitem [{\citenamefont {Guo}\ \emph {et~al.}(2019)\citenamefont {Guo},
  \citenamefont {Zhang},\ and\ \citenamefont {Zhang}}]{Guo:2018ans}%
  \BibitemOpen
  \bibfield  {author} {\bibinfo {author} {\bibfnamefont {R.-Y.}\ \bibnamefont
  {Guo}}, \bibinfo {author} {\bibfnamefont {J.-F.}\ \bibnamefont {Zhang}}, \
  and\ \bibinfo {author} {\bibfnamefont {X.}~\bibnamefont {Zhang}},\ }\href
  {\doibase 10.1088/1475-7516/2019/02/054} {\bibfield  {journal} {\bibinfo
  {journal} {JCAP}\ }\textbf {\bibinfo {volume} {1902}},\ \bibinfo {pages}
  {054} (\bibinfo {year} {2019})},\ \Eprint {http://arxiv.org/abs/1809.02340}
  {arXiv:1809.02340 [astro-ph.CO]} \BibitemShut {NoStop}%
\bibitem [{\citenamefont {Kumar}\ \emph {et~al.}(2019)\citenamefont {Kumar},
  \citenamefont {Nunes},\ and\ \citenamefont {Yadav}}]{Kumar:2019wfs}%
  \BibitemOpen
  \bibfield  {author} {\bibinfo {author} {\bibfnamefont {S.}~\bibnamefont
  {Kumar}}, \bibinfo {author} {\bibfnamefont {R.~C.}\ \bibnamefont {Nunes}}, \
  and\ \bibinfo {author} {\bibfnamefont {S.~K.}\ \bibnamefont {Yadav}},\
  }\href@noop {} {\  (\bibinfo {year} {2019})},\ \Eprint
  {http://arxiv.org/abs/1903.04865} {arXiv:1903.04865 [astro-ph.CO]}
  \BibitemShut {NoStop}%
\bibitem [{\citenamefont {Miao}\ and\ \citenamefont
  {Huang}(2018)}]{Miao:2018zpw}%
  \BibitemOpen
  \bibfield  {author} {\bibinfo {author} {\bibfnamefont {H.}~\bibnamefont
  {Miao}}\ and\ \bibinfo {author} {\bibfnamefont {Z.}~\bibnamefont {Huang}},\
  }\href {\doibase 10.3847/1538-4357/aae523} {\bibfield  {journal} {\bibinfo
  {journal} {Astrophys. J.}\ }\textbf {\bibinfo {volume} {868}},\ \bibinfo
  {pages} {20} (\bibinfo {year} {2018})},\ \Eprint
  {http://arxiv.org/abs/1803.07320} {arXiv:1803.07320 [astro-ph.CO]}
  \BibitemShut {NoStop}%
\bibitem [{\citenamefont {Di~Valentino}\ \emph
  {et~al.}(2019{\natexlab{a}})\citenamefont {Di~Valentino}, \citenamefont
  {Ferreira}, \citenamefont {Visinelli},\ and\ \citenamefont
  {Danielsson}}]{DiValentino:2019exe}%
  \BibitemOpen
  \bibfield  {author} {\bibinfo {author} {\bibfnamefont {E.}~\bibnamefont
  {Di~Valentino}}, \bibinfo {author} {\bibfnamefont {R.~Z.}\ \bibnamefont
  {Ferreira}}, \bibinfo {author} {\bibfnamefont {L.}~\bibnamefont {Visinelli}},
  \ and\ \bibinfo {author} {\bibfnamefont {U.}~\bibnamefont {Danielsson}},\
  }\href {\doibase 10.1016/j.dark.2019.100385} {\bibfield  {journal} {\bibinfo
  {journal} {Phys. Dark Univ.}\ }\textbf {\bibinfo {volume} {26}},\ \bibinfo
  {pages} {100385} (\bibinfo {year} {2019}{\natexlab{a}})},\ \Eprint
  {http://arxiv.org/abs/1906.11255} {arXiv:1906.11255 [astro-ph.CO]}
  \BibitemShut {NoStop}%
\bibitem [{\citenamefont {Ko}\ \emph {et~al.}(2017)\citenamefont {Ko},
  \citenamefont {Nagata},\ and\ \citenamefont {Tang}}]{Ko:2017uyb}%
  \BibitemOpen
  \bibfield  {author} {\bibinfo {author} {\bibfnamefont {P.}~\bibnamefont
  {Ko}}, \bibinfo {author} {\bibfnamefont {N.}~\bibnamefont {Nagata}}, \ and\
  \bibinfo {author} {\bibfnamefont {Y.}~\bibnamefont {Tang}},\ }\href {\doibase
  10.1016/j.physletb.2017.08.065} {\bibfield  {journal} {\bibinfo  {journal}
  {Phys. Lett.}\ }\textbf {\bibinfo {volume} {B773}},\ \bibinfo {pages} {513}
  (\bibinfo {year} {2017})},\ \Eprint {http://arxiv.org/abs/1706.05605}
  {arXiv:1706.05605 [hep-ph]} \BibitemShut {NoStop}%
\bibitem [{\citenamefont {Raveri}\ \emph {et~al.}(2017)\citenamefont {Raveri},
  \citenamefont {Hu}, \citenamefont {Hoffman},\ and\ \citenamefont
  {Wang}}]{Raveri:2017jto}%
  \BibitemOpen
  \bibfield  {author} {\bibinfo {author} {\bibfnamefont {M.}~\bibnamefont
  {Raveri}}, \bibinfo {author} {\bibfnamefont {W.}~\bibnamefont {Hu}}, \bibinfo
  {author} {\bibfnamefont {T.}~\bibnamefont {Hoffman}}, \ and\ \bibinfo
  {author} {\bibfnamefont {L.-T.}\ \bibnamefont {Wang}},\ }\href {\doibase
  10.1103/PhysRevD.96.103501} {\bibfield  {journal} {\bibinfo  {journal} {Phys.
  Rev.}\ }\textbf {\bibinfo {volume} {D96}},\ \bibinfo {pages} {103501}
  (\bibinfo {year} {2017})},\ \Eprint {http://arxiv.org/abs/1709.04877}
  {arXiv:1709.04877 [astro-ph.CO]} \BibitemShut {NoStop}%
\bibitem [{\citenamefont {Buen-Abad}\ \emph {et~al.}(2018)\citenamefont
  {Buen-Abad}, \citenamefont {Emami},\ and\ \citenamefont
  {Schmaltz}}]{Buen-Abad:2018mas}%
  \BibitemOpen
  \bibfield  {author} {\bibinfo {author} {\bibfnamefont {M.~A.}\ \bibnamefont
  {Buen-Abad}}, \bibinfo {author} {\bibfnamefont {R.}~\bibnamefont {Emami}}, \
  and\ \bibinfo {author} {\bibfnamefont {M.}~\bibnamefont {Schmaltz}},\ }\href
  {\doibase 10.1103/PhysRevD.98.083517} {\bibfield  {journal} {\bibinfo
  {journal} {Phys. Rev.}\ }\textbf {\bibinfo {volume} {D98}},\ \bibinfo {pages}
  {083517} (\bibinfo {year} {2018})},\ \Eprint
  {http://arxiv.org/abs/1803.08062} {arXiv:1803.08062 [hep-ph]} \BibitemShut
  {NoStop}%
\bibitem [{\citenamefont {Choi}\ \emph {et~al.}(2019)\citenamefont {Choi},
  \citenamefont {Suzuki},\ and\ \citenamefont {Yanagida}}]{Choi:2019jck}%
  \BibitemOpen
  \bibfield  {author} {\bibinfo {author} {\bibfnamefont {G.}~\bibnamefont
  {Choi}}, \bibinfo {author} {\bibfnamefont {M.}~\bibnamefont {Suzuki}}, \ and\
  \bibinfo {author} {\bibfnamefont {T.~T.}\ \bibnamefont {Yanagida}},\
  }\href@noop {} {\  (\bibinfo {year} {2019})},\ \Eprint
  {http://arxiv.org/abs/1910.00459} {arXiv:1910.00459 [hep-ph]} \BibitemShut
  {NoStop}%
\bibitem [{\citenamefont {D'Eramo}\ \emph {et~al.}(2018)\citenamefont
  {D'Eramo}, \citenamefont {Ferreira}, \citenamefont {Notari},\ and\
  \citenamefont {Bernal}}]{DEramo:2018vss}%
  \BibitemOpen
  \bibfield  {author} {\bibinfo {author} {\bibfnamefont {F.}~\bibnamefont
  {D'Eramo}}, \bibinfo {author} {\bibfnamefont {R.~Z.}\ \bibnamefont
  {Ferreira}}, \bibinfo {author} {\bibfnamefont {A.}~\bibnamefont {Notari}}, \
  and\ \bibinfo {author} {\bibfnamefont {J.~L.}\ \bibnamefont {Bernal}},\
  }\href {\doibase 10.1088/1475-7516/2018/11/014} {\bibfield  {journal}
  {\bibinfo  {journal} {JCAP}\ }\textbf {\bibinfo {volume} {1811}},\ \bibinfo
  {pages} {014} (\bibinfo {year} {2018})},\ \Eprint
  {http://arxiv.org/abs/1808.07430} {arXiv:1808.07430 [hep-ph]} \BibitemShut
  {NoStop}%
\bibitem [{\citenamefont {Yang}\ \emph
  {et~al.}(2019{\natexlab{a}})\citenamefont {Yang}, \citenamefont {Pan},
  \citenamefont {Xu},\ and\ \citenamefont {Mota}}]{Yang:2018ubt}%
  \BibitemOpen
  \bibfield  {author} {\bibinfo {author} {\bibfnamefont {W.}~\bibnamefont
  {Yang}}, \bibinfo {author} {\bibfnamefont {S.}~\bibnamefont {Pan}}, \bibinfo
  {author} {\bibfnamefont {L.}~\bibnamefont {Xu}}, \ and\ \bibinfo {author}
  {\bibfnamefont {D.~F.}\ \bibnamefont {Mota}},\ }\href {\doibase
  10.1093/mnras/sty2789} {\bibfield  {journal} {\bibinfo  {journal} {Mon. Not.
  Roy. Astron. Soc.}\ }\textbf {\bibinfo {volume} {482}},\ \bibinfo {pages}
  {1858} (\bibinfo {year} {2019}{\natexlab{a}})},\ \Eprint
  {http://arxiv.org/abs/1804.08455} {arXiv:1804.08455 [astro-ph.CO]}
  \BibitemShut {NoStop}%
\bibitem [{\citenamefont {Yang}\ \emph
  {et~al.}(2019{\natexlab{b}})\citenamefont {Yang}, \citenamefont {Pan},\ and\
  \citenamefont {Paliathanasis}}]{Yang:2018pej}%
  \BibitemOpen
  \bibfield  {author} {\bibinfo {author} {\bibfnamefont {W.}~\bibnamefont
  {Yang}}, \bibinfo {author} {\bibfnamefont {S.}~\bibnamefont {Pan}}, \ and\
  \bibinfo {author} {\bibfnamefont {A.}~\bibnamefont {Paliathanasis}},\ }\href
  {\doibase 10.1093/mnras/sty2780} {\bibfield  {journal} {\bibinfo  {journal}
  {Mon. Not. Roy. Astron. Soc.}\ }\textbf {\bibinfo {volume} {482}},\ \bibinfo
  {pages} {1007} (\bibinfo {year} {2019}{\natexlab{b}})},\ \Eprint
  {http://arxiv.org/abs/1804.08558} {arXiv:1804.08558 [gr-qc]} \BibitemShut
  {NoStop}%
\bibitem [{\citenamefont {Yang}\ \emph {et~al.}(2018)\citenamefont {Yang},
  \citenamefont {Pan}, \citenamefont {Di~Valentino}, \citenamefont {Nunes},
  \citenamefont {Vagnozzi},\ and\ \citenamefont {Mota}}]{Yang:2018euj}%
  \BibitemOpen
  \bibfield  {author} {\bibinfo {author} {\bibfnamefont {W.}~\bibnamefont
  {Yang}}, \bibinfo {author} {\bibfnamefont {S.}~\bibnamefont {Pan}}, \bibinfo
  {author} {\bibfnamefont {E.}~\bibnamefont {Di~Valentino}}, \bibinfo {author}
  {\bibfnamefont {R.~C.}\ \bibnamefont {Nunes}}, \bibinfo {author}
  {\bibfnamefont {S.}~\bibnamefont {Vagnozzi}}, \ and\ \bibinfo {author}
  {\bibfnamefont {D.~F.}\ \bibnamefont {Mota}},\ }\href {\doibase
  10.1088/1475-7516/2018/09/019} {\bibfield  {journal} {\bibinfo  {journal}
  {JCAP}\ }\textbf {\bibinfo {volume} {1809}},\ \bibinfo {pages} {019}
  (\bibinfo {year} {2018})},\ \Eprint {http://arxiv.org/abs/1805.08252}
  {arXiv:1805.08252 [astro-ph.CO]} \BibitemShut {NoStop}%
\bibitem [{\citenamefont {Blinov}\ \emph {et~al.}(2019)\citenamefont {Blinov},
  \citenamefont {Kelly}, \citenamefont {Krnjaic},\ and\ \citenamefont
  {McDermott}}]{Blinov:2019gcj}%
  \BibitemOpen
  \bibfield  {author} {\bibinfo {author} {\bibfnamefont {N.}~\bibnamefont
  {Blinov}}, \bibinfo {author} {\bibfnamefont {K.~J.}\ \bibnamefont {Kelly}},
  \bibinfo {author} {\bibfnamefont {G.~Z.}\ \bibnamefont {Krnjaic}}, \ and\
  \bibinfo {author} {\bibfnamefont {S.~D.}\ \bibnamefont {McDermott}},\
  }\href@noop {} {\  (\bibinfo {year} {2019})},\ \Eprint
  {http://arxiv.org/abs/1905.02727} {arXiv:1905.02727 [astro-ph.CO]}
  \BibitemShut {NoStop}%
\bibitem [{\citenamefont {Escudero}\ \emph {et~al.}(2019)\citenamefont
  {Escudero}, \citenamefont {Hooper}, \citenamefont {Krnjaic},\ and\
  \citenamefont {Pierre}}]{Escudero:2019gzq}%
  \BibitemOpen
  \bibfield  {author} {\bibinfo {author} {\bibfnamefont {M.}~\bibnamefont
  {Escudero}}, \bibinfo {author} {\bibfnamefont {D.}~\bibnamefont {Hooper}},
  \bibinfo {author} {\bibfnamefont {G.}~\bibnamefont {Krnjaic}}, \ and\
  \bibinfo {author} {\bibfnamefont {M.}~\bibnamefont {Pierre}},\ }\href
  {\doibase 10.1007/JHEP03(2019)071} {\bibfield  {journal} {\bibinfo  {journal}
  {JHEP}\ }\textbf {\bibinfo {volume} {03}},\ \bibinfo {pages} {071} (\bibinfo
  {year} {2019})},\ \Eprint {http://arxiv.org/abs/1901.02010} {arXiv:1901.02010
  [hep-ph]} \BibitemShut {NoStop}%
\bibitem [{\citenamefont {Anchordoqui}\ \emph {et~al.}(2015)\citenamefont
  {Anchordoqui}, \citenamefont {Barger}, \citenamefont {Goldberg},
  \citenamefont {Huang}, \citenamefont {Marfatia}, \citenamefont {da~Silva},\
  and\ \citenamefont {Weiler}}]{Anchordoqui:2015lqa}%
  \BibitemOpen
  \bibfield  {author} {\bibinfo {author} {\bibfnamefont {L.~A.}\ \bibnamefont
  {Anchordoqui}}, \bibinfo {author} {\bibfnamefont {V.}~\bibnamefont {Barger}},
  \bibinfo {author} {\bibfnamefont {H.}~\bibnamefont {Goldberg}}, \bibinfo
  {author} {\bibfnamefont {X.}~\bibnamefont {Huang}}, \bibinfo {author}
  {\bibfnamefont {D.}~\bibnamefont {Marfatia}}, \bibinfo {author}
  {\bibfnamefont {L.~H.~M.}\ \bibnamefont {da~Silva}}, \ and\ \bibinfo {author}
  {\bibfnamefont {T.~J.}\ \bibnamefont {Weiler}},\ }\href {\doibase
  10.1103/PhysRevD.92.061301, 10.1103/PhysRevD.94.069901} {\bibfield  {journal}
  {\bibinfo  {journal} {Phys. Rev.}\ }\textbf {\bibinfo {volume} {D92}},\
  \bibinfo {pages} {061301} (\bibinfo {year} {2015})},\ \bibinfo {note}
  {[Erratum: Phys. Rev.D94,no.6,069901(2016)]},\ \Eprint
  {http://arxiv.org/abs/1506.08788} {arXiv:1506.08788 [hep-ph]} \BibitemShut
  {NoStop}%
\bibitem [{\citenamefont {Buch}\ \emph {et~al.}(2017)\citenamefont {Buch},
  \citenamefont {Ralegankar},\ and\ \citenamefont {Rentala}}]{Buch:2016jjp}%
  \BibitemOpen
  \bibfield  {author} {\bibinfo {author} {\bibfnamefont {J.}~\bibnamefont
  {Buch}}, \bibinfo {author} {\bibfnamefont {P.}~\bibnamefont {Ralegankar}}, \
  and\ \bibinfo {author} {\bibfnamefont {V.}~\bibnamefont {Rentala}},\ }\href
  {\doibase 10.1088/1475-7516/2017/10/028} {\bibfield  {journal} {\bibinfo
  {journal} {JCAP}\ }\textbf {\bibinfo {volume} {1710}},\ \bibinfo {pages}
  {028} (\bibinfo {year} {2017})},\ \Eprint {http://arxiv.org/abs/1609.04821}
  {arXiv:1609.04821 [hep-ph]} \BibitemShut {NoStop}%
\bibitem [{\citenamefont {Bringmann}\ \emph {et~al.}(2018)\citenamefont
  {Bringmann}, \citenamefont {Kahlhoefer}, \citenamefont {Schmidt-Hoberg},\
  and\ \citenamefont {Walia}}]{Bringmann:2018jpr}%
  \BibitemOpen
  \bibfield  {author} {\bibinfo {author} {\bibfnamefont {T.}~\bibnamefont
  {Bringmann}}, \bibinfo {author} {\bibfnamefont {F.}~\bibnamefont
  {Kahlhoefer}}, \bibinfo {author} {\bibfnamefont {K.}~\bibnamefont
  {Schmidt-Hoberg}}, \ and\ \bibinfo {author} {\bibfnamefont {P.}~\bibnamefont
  {Walia}},\ }\href {\doibase 10.1103/PhysRevD.98.023543} {\bibfield  {journal}
  {\bibinfo  {journal} {Phys. Rev.}\ }\textbf {\bibinfo {volume} {D98}},\
  \bibinfo {pages} {023543} (\bibinfo {year} {2018})},\ \Eprint
  {http://arxiv.org/abs/1803.03644} {arXiv:1803.03644 [astro-ph.CO]}
  \BibitemShut {NoStop}%
\bibitem [{\citenamefont {Pandey}\ \emph {et~al.}(2019)\citenamefont {Pandey},
  \citenamefont {Karwal},\ and\ \citenamefont {Das}}]{Pandey:2019plg}%
  \BibitemOpen
  \bibfield  {author} {\bibinfo {author} {\bibfnamefont {K.~L.}\ \bibnamefont
  {Pandey}}, \bibinfo {author} {\bibfnamefont {T.}~\bibnamefont {Karwal}}, \
  and\ \bibinfo {author} {\bibfnamefont {S.}~\bibnamefont {Das}},\ }\href@noop
  {} {\  (\bibinfo {year} {2019})},\ \Eprint {http://arxiv.org/abs/1902.10636}
  {arXiv:1902.10636 [astro-ph.CO]} \BibitemShut {NoStop}%
\bibitem [{\citenamefont {{Doroshkevich}}\ and\ \citenamefont
  {{Khlopov}}(1985)}]{Doroshkevich:1985}%
  \BibitemOpen
  \bibfield  {author} {\bibinfo {author} {\bibfnamefont {A.~G.}\ \bibnamefont
  {{Doroshkevich}}}\ and\ \bibinfo {author} {\bibfnamefont {M.~Y.}\
  \bibnamefont {{Khlopov}}},\ }\href@noop {} {\bibfield  {journal} {\bibinfo
  {journal} {Soviet Astronomy Letters}\ }\textbf {\bibinfo {volume} {11}},\
  \bibinfo {pages} {236} (\bibinfo {year} {1985})}\BibitemShut {NoStop}%
\bibitem [{\citenamefont {{Doroshkevich}}\ and\ \citenamefont
  {{Khlopov}}(1984)}]{Doroshkevich:1984ster}%
  \BibitemOpen
  \bibfield  {author} {\bibinfo {author} {\bibfnamefont {A.~G.}\ \bibnamefont
  {{Doroshkevich}}}\ and\ \bibinfo {author} {\bibfnamefont {M.~I.}\
  \bibnamefont {{Khlopov}}},\ }\href {\doibase 10.1093/mnras/211.2.277}
  {\bibfield  {journal} {\bibinfo  {journal} {Mon. Not. Roy. Astron. Soc.}\
  }\textbf {\bibinfo {volume} {211}},\ \bibinfo {pages} {277} (\bibinfo {year}
  {1984})}\BibitemShut {NoStop}%
\bibitem [{\citenamefont {Vattis}\ \emph {et~al.}(2019)\citenamefont {Vattis},
  \citenamefont {Koushiappas},\ and\ \citenamefont {Loeb}}]{Vattis:2019efj}%
  \BibitemOpen
  \bibfield  {author} {\bibinfo {author} {\bibfnamefont {K.}~\bibnamefont
  {Vattis}}, \bibinfo {author} {\bibfnamefont {S.~M.}\ \bibnamefont
  {Koushiappas}}, \ and\ \bibinfo {author} {\bibfnamefont {A.}~\bibnamefont
  {Loeb}},\ }\href@noop {} {\  (\bibinfo {year} {2019})},\ \Eprint
  {http://arxiv.org/abs/1903.06220} {arXiv:1903.06220 [astro-ph.CO]}
  \BibitemShut {NoStop}%
\bibitem [{\citenamefont {Di~Valentino}\ \emph
  {et~al.}(2019{\natexlab{b}})\citenamefont {Di~Valentino}, \citenamefont
  {Melchiorri}, \citenamefont {Mena},\ and\ \citenamefont
  {Vagnozzi}}]{DiValentino:2019ffd}%
  \BibitemOpen
  \bibfield  {author} {\bibinfo {author} {\bibfnamefont {E.}~\bibnamefont
  {Di~Valentino}}, \bibinfo {author} {\bibfnamefont {A.}~\bibnamefont
  {Melchiorri}}, \bibinfo {author} {\bibfnamefont {O.}~\bibnamefont {Mena}}, \
  and\ \bibinfo {author} {\bibfnamefont {S.}~\bibnamefont {Vagnozzi}},\
  }\href@noop {} {\  (\bibinfo {year} {2019}{\natexlab{b}})},\ \Eprint
  {http://arxiv.org/abs/1908.04281} {arXiv:1908.04281 [astro-ph.CO]}
  \BibitemShut {NoStop}%
\bibitem [{\citenamefont {Pan}\ \emph {et~al.}(2019)\citenamefont {Pan},
  \citenamefont {Yang}, \citenamefont {Di~Valentino}, \citenamefont
  {Saridakis},\ and\ \citenamefont {Chakraborty}}]{Pan:2019gop}%
  \BibitemOpen
  \bibfield  {author} {\bibinfo {author} {\bibfnamefont {S.}~\bibnamefont
  {Pan}}, \bibinfo {author} {\bibfnamefont {W.}~\bibnamefont {Yang}}, \bibinfo
  {author} {\bibfnamefont {E.}~\bibnamefont {Di~Valentino}}, \bibinfo {author}
  {\bibfnamefont {E.~N.}\ \bibnamefont {Saridakis}}, \ and\ \bibinfo {author}
  {\bibfnamefont {S.}~\bibnamefont {Chakraborty}},\ }\href@noop {} {\
  (\bibinfo {year} {2019})},\ \Eprint {http://arxiv.org/abs/1907.07540}
  {arXiv:1907.07540 [astro-ph.CO]} \BibitemShut {NoStop}%
\bibitem [{\citenamefont {Vagnozzi}(2019)}]{Vagnozzi:2019ezj}%
  \BibitemOpen
  \bibfield  {author} {\bibinfo {author} {\bibfnamefont {S.}~\bibnamefont
  {Vagnozzi}},\ }\href@noop {} {\  (\bibinfo {year} {2019})},\ \Eprint
  {http://arxiv.org/abs/1907.07569} {arXiv:1907.07569 [astro-ph.CO]}
  \BibitemShut {NoStop}%
\bibitem [{\citenamefont {Battye}\ and\ \citenamefont
  {Moss}(2014)}]{Battye:2013xqa}%
  \BibitemOpen
  \bibfield  {author} {\bibinfo {author} {\bibfnamefont {R.~A.}\ \bibnamefont
  {Battye}}\ and\ \bibinfo {author} {\bibfnamefont {A.}~\bibnamefont {Moss}},\
  }\href {\doibase 10.1103/PhysRevLett.112.051303} {\bibfield  {journal}
  {\bibinfo  {journal} {Phys. Rev. Lett.}\ }\textbf {\bibinfo {volume} {112}},\
  \bibinfo {pages} {051303} (\bibinfo {year} {2014})},\ \Eprint
  {http://arxiv.org/abs/1308.5870} {arXiv:1308.5870 [astro-ph.CO]} \BibitemShut
  {NoStop}%
\bibitem [{\citenamefont {Arias-Aragon}\ \emph {et~al.}(2020)\citenamefont
  {Arias-Aragon}, \citenamefont {Fernandez-Martinez}, \citenamefont
  {Gonzalez-Lopez},\ and\ \citenamefont {Merlo}}]{Arias-Aragon:2020qip}%
  \BibitemOpen
  \bibfield  {author} {\bibinfo {author} {\bibfnamefont {F.}~\bibnamefont
  {Arias-Aragon}}, \bibinfo {author} {\bibfnamefont {E.}~\bibnamefont
  {Fernandez-Martinez}}, \bibinfo {author} {\bibfnamefont {M.}~\bibnamefont
  {Gonzalez-Lopez}}, \ and\ \bibinfo {author} {\bibfnamefont {L.}~\bibnamefont
  {Merlo}},\ }\href@noop {} {\  (\bibinfo {year} {2020})},\ \Eprint
  {http://arxiv.org/abs/2009.01848} {arXiv:2009.01848 [hep-ph]} \BibitemShut
  {NoStop}%
\bibitem [{\citenamefont {Flores}\ and\ \citenamefont
  {Kusenko}(2020)}]{Flores:2020drq}%
  \BibitemOpen
  \bibfield  {author} {\bibinfo {author} {\bibfnamefont {M.~M.}\ \bibnamefont
  {Flores}}\ and\ \bibinfo {author} {\bibfnamefont {A.}~\bibnamefont
  {Kusenko}},\ }\href@noop {} {\  (\bibinfo {year} {2020})},\ \Eprint
  {http://arxiv.org/abs/2008.12456} {arXiv:2008.12456 [astro-ph.CO]}
  \BibitemShut {NoStop}%
\bibitem [{\citenamefont {Bernal}\ \emph {et~al.}(2016)\citenamefont {Bernal},
  \citenamefont {Verde},\ and\ \citenamefont {Riess}}]{Bernal:2016gxb}%
  \BibitemOpen
  \bibfield  {author} {\bibinfo {author} {\bibfnamefont {J.~L.}\ \bibnamefont
  {Bernal}}, \bibinfo {author} {\bibfnamefont {L.}~\bibnamefont {Verde}}, \
  and\ \bibinfo {author} {\bibfnamefont {A.~G.}\ \bibnamefont {Riess}},\ }\href
  {\doibase 10.1088/1475-7516/2016/10/019} {\bibfield  {journal} {\bibinfo
  {journal} {JCAP}\ }\textbf {\bibinfo {volume} {1610}},\ \bibinfo {pages}
  {019} (\bibinfo {year} {2016})},\ \Eprint {http://arxiv.org/abs/1607.05617}
  {arXiv:1607.05617 [astro-ph.CO]} \BibitemShut {NoStop}%
\bibitem [{\citenamefont {Mangano}\ \emph {et~al.}(2005)\citenamefont
  {Mangano}, \citenamefont {Miele}, \citenamefont {Pastor}, \citenamefont
  {Pinto}, \citenamefont {Pisanti},\ and\ \citenamefont
  {Serpico}}]{Mangano:2005cc}%
  \BibitemOpen
  \bibfield  {author} {\bibinfo {author} {\bibfnamefont {G.}~\bibnamefont
  {Mangano}}, \bibinfo {author} {\bibfnamefont {G.}~\bibnamefont {Miele}},
  \bibinfo {author} {\bibfnamefont {S.}~\bibnamefont {Pastor}}, \bibinfo
  {author} {\bibfnamefont {T.}~\bibnamefont {Pinto}}, \bibinfo {author}
  {\bibfnamefont {O.}~\bibnamefont {Pisanti}}, \ and\ \bibinfo {author}
  {\bibfnamefont {P.~D.}\ \bibnamefont {Serpico}},\ }\href {\doibase
  10.1016/j.nuclphysb.2005.09.041} {\bibfield  {journal} {\bibinfo  {journal}
  {Nucl. Phys.}\ }\textbf {\bibinfo {volume} {B729}},\ \bibinfo {pages} {221}
  (\bibinfo {year} {2005})},\ \Eprint {http://arxiv.org/abs/hep-ph/0506164}
  {arXiv:hep-ph/0506164 [hep-ph]} \BibitemShut {NoStop}%
\bibitem [{\citenamefont {Kersten}\ and\ \citenamefont
  {Smirnov}(2007)}]{Kersten:2007vk}%
  \BibitemOpen
  \bibfield  {author} {\bibinfo {author} {\bibfnamefont {J.}~\bibnamefont
  {Kersten}}\ and\ \bibinfo {author} {\bibfnamefont {A.~Y.}\ \bibnamefont
  {Smirnov}},\ }\href {\doibase 10.1103/PhysRevD.76.073005} {\bibfield
  {journal} {\bibinfo  {journal} {Phys. Rev. D}\ }\textbf {\bibinfo {volume}
  {76}},\ \bibinfo {pages} {073005} (\bibinfo {year} {2007})},\ \Eprint
  {http://arxiv.org/abs/0705.3221} {arXiv:0705.3221 [hep-ph]} \BibitemShut
  {NoStop}%
\bibitem [{\citenamefont {Drewes}\ \emph {et~al.}(2018)\citenamefont {Drewes},
  \citenamefont {Hajer}, \citenamefont {Klaric},\ and\ \citenamefont
  {Lanfranchi}}]{Drewes:2018gkc}%
  \BibitemOpen
  \bibfield  {author} {\bibinfo {author} {\bibfnamefont {M.}~\bibnamefont
  {Drewes}}, \bibinfo {author} {\bibfnamefont {J.}~\bibnamefont {Hajer}},
  \bibinfo {author} {\bibfnamefont {J.}~\bibnamefont {Klaric}}, \ and\ \bibinfo
  {author} {\bibfnamefont {G.}~\bibnamefont {Lanfranchi}},\ }\href {\doibase
  10.1007/JHEP07(2018)105} {\bibfield  {journal} {\bibinfo  {journal} {JHEP}\
  }\textbf {\bibinfo {volume} {07}},\ \bibinfo {pages} {105} (\bibinfo {year}
  {2018})},\ \Eprint {http://arxiv.org/abs/1801.04207} {arXiv:1801.04207
  [hep-ph]} \BibitemShut {NoStop}%
\bibitem [{\citenamefont {Chrzaszcz}\ \emph {et~al.}(2020)\citenamefont
  {Chrzaszcz}, \citenamefont {Drewes}, \citenamefont {Gonzalo}, \citenamefont
  {Harz}, \citenamefont {Krishnamurthy},\ and\ \citenamefont
  {Weniger}}]{Chrzaszcz:2019inj}%
  \BibitemOpen
  \bibfield  {author} {\bibinfo {author} {\bibfnamefont {M.}~\bibnamefont
  {Chrzaszcz}}, \bibinfo {author} {\bibfnamefont {M.}~\bibnamefont {Drewes}},
  \bibinfo {author} {\bibfnamefont {T.~E.}\ \bibnamefont {Gonzalo}}, \bibinfo
  {author} {\bibfnamefont {J.}~\bibnamefont {Harz}}, \bibinfo {author}
  {\bibfnamefont {S.}~\bibnamefont {Krishnamurthy}}, \ and\ \bibinfo {author}
  {\bibfnamefont {C.}~\bibnamefont {Weniger}},\ }\href {\doibase
  10.1140/epjc/s10052-020-8073-9} {\bibfield  {journal} {\bibinfo  {journal}
  {Eur. Phys. J. C}\ }\textbf {\bibinfo {volume} {80}},\ \bibinfo {pages} {569}
  (\bibinfo {year} {2020})},\ \Eprint {http://arxiv.org/abs/1908.02302}
  {arXiv:1908.02302 [hep-ph]} \BibitemShut {NoStop}%
\bibitem [{\citenamefont {Sabti}\ \emph {et~al.}(2020)\citenamefont {Sabti},
  \citenamefont {Magalich},\ and\ \citenamefont {Filimonova}}]{Sabti:2020yrt}%
  \BibitemOpen
  \bibfield  {author} {\bibinfo {author} {\bibfnamefont {N.}~\bibnamefont
  {Sabti}}, \bibinfo {author} {\bibfnamefont {A.}~\bibnamefont {Magalich}}, \
  and\ \bibinfo {author} {\bibfnamefont {A.}~\bibnamefont {Filimonova}},\
  }\href {\doibase 10.1088/1475-7516/2020/11/056} {\bibfield  {journal}
  {\bibinfo  {journal} {JCAP}\ }\textbf {\bibinfo {volume} {11}},\ \bibinfo
  {pages} {056} (\bibinfo {year} {2020})},\ \Eprint
  {http://arxiv.org/abs/2006.07387} {arXiv:2006.07387 [hep-ph]} \BibitemShut
  {NoStop}%
\bibitem [{\citenamefont {Aoki}\ \emph {et~al.}(2011)\citenamefont {Aoki} \emph
  {et~al.}}]{PIENU:2011aa}%
  \BibitemOpen
  \bibfield  {author} {\bibinfo {author} {\bibfnamefont {M.}~\bibnamefont
  {Aoki}} \emph {et~al.} (\bibinfo {collaboration} {PIENU}),\ }\href {\doibase
  10.1103/PhysRevD.84.052002} {\bibfield  {journal} {\bibinfo  {journal} {Phys.
  Rev.}\ }\textbf {\bibinfo {volume} {D84}},\ \bibinfo {pages} {052002}
  (\bibinfo {year} {2011})},\ \Eprint {http://arxiv.org/abs/1106.4055}
  {arXiv:1106.4055 [hep-ex]} \BibitemShut {NoStop}%
\bibitem [{\citenamefont {Britton}\ \emph
  {et~al.}(1992{\natexlab{a}})\citenamefont {Britton}, \citenamefont {Ahmad},
  \citenamefont {Bryman}, \citenamefont {Burnham}, \citenamefont {Clifford},
  \citenamefont {Kitching}, \citenamefont {Kuno}, \citenamefont {Macdonald},
  \citenamefont {Numao}, \citenamefont {Olin}, \citenamefont {Poutissou},\ and\
  \citenamefont {Dixit}}]{Britton:1992}%
  \BibitemOpen
  \bibfield  {author} {\bibinfo {author} {\bibfnamefont {D.~I.}\ \bibnamefont
  {Britton}}, \bibinfo {author} {\bibfnamefont {S.}~\bibnamefont {Ahmad}},
  \bibinfo {author} {\bibfnamefont {D.~A.}\ \bibnamefont {Bryman}}, \bibinfo
  {author} {\bibfnamefont {R.~A.}\ \bibnamefont {Burnham}}, \bibinfo {author}
  {\bibfnamefont {E.~T.~H.}\ \bibnamefont {Clifford}}, \bibinfo {author}
  {\bibfnamefont {P.}~\bibnamefont {Kitching}}, \bibinfo {author}
  {\bibfnamefont {Y.}~\bibnamefont {Kuno}}, \bibinfo {author} {\bibfnamefont
  {J.~A.}\ \bibnamefont {Macdonald}}, \bibinfo {author} {\bibfnamefont
  {T.}~\bibnamefont {Numao}}, \bibinfo {author} {\bibfnamefont
  {A.}~\bibnamefont {Olin}}, \bibinfo {author} {\bibfnamefont {J.~M.}\
  \bibnamefont {Poutissou}}, \ and\ \bibinfo {author} {\bibfnamefont {M.~S.}\
  \bibnamefont {Dixit}},\ }\href {\doibase 10.1103/PhysRevD.46.R885} {\bibfield
   {journal} {\bibinfo  {journal} {Phys. Rev. D}\ }\textbf {\bibinfo {volume}
  {46}},\ \bibinfo {pages} {R885} (\bibinfo {year}
  {1992}{\natexlab{a}})}\BibitemShut {NoStop}%
\bibitem [{\citenamefont {Britton}\ \emph
  {et~al.}(1992{\natexlab{b}})\citenamefont {Britton}, \citenamefont {Ahmad},
  \citenamefont {Bryman}, \citenamefont {Burnham}, \citenamefont {Clifford},
  \citenamefont {Kitching}, \citenamefont {Kuno}, \citenamefont {Macdonald},
  \citenamefont {Numao}, \citenamefont {Olin}, \citenamefont {Poutissou},\ and\
  \citenamefont {Dixit}}]{Britton:1992prl}%
  \BibitemOpen
  \bibfield  {author} {\bibinfo {author} {\bibfnamefont {D.~I.}\ \bibnamefont
  {Britton}}, \bibinfo {author} {\bibfnamefont {S.}~\bibnamefont {Ahmad}},
  \bibinfo {author} {\bibfnamefont {D.~A.}\ \bibnamefont {Bryman}}, \bibinfo
  {author} {\bibfnamefont {R.~A.}\ \bibnamefont {Burnham}}, \bibinfo {author}
  {\bibfnamefont {E.~T.~H.}\ \bibnamefont {Clifford}}, \bibinfo {author}
  {\bibfnamefont {P.}~\bibnamefont {Kitching}}, \bibinfo {author}
  {\bibfnamefont {Y.}~\bibnamefont {Kuno}}, \bibinfo {author} {\bibfnamefont
  {J.~A.}\ \bibnamefont {Macdonald}}, \bibinfo {author} {\bibfnamefont
  {T.}~\bibnamefont {Numao}}, \bibinfo {author} {\bibfnamefont
  {A.}~\bibnamefont {Olin}}, \bibinfo {author} {\bibfnamefont {J.-M.}\
  \bibnamefont {Poutissou}}, \ and\ \bibinfo {author} {\bibfnamefont {M.~S.}\
  \bibnamefont {Dixit}},\ }\href {\doibase 10.1103/PhysRevLett.68.3000}
  {\bibfield  {journal} {\bibinfo  {journal} {Phys. Rev. Lett.}\ }\textbf
  {\bibinfo {volume} {68}},\ \bibinfo {pages} {3000} (\bibinfo {year}
  {1992}{\natexlab{b}})}\BibitemShut {NoStop}%
\bibitem [{\citenamefont {Aguilar-Arevalo}\ \emph {et~al.}(2019)\citenamefont
  {Aguilar-Arevalo} \emph {et~al.}}]{Aguilar-Arevalo:2019owf}%
  \BibitemOpen
  \bibfield  {author} {\bibinfo {author} {\bibfnamefont {A.}~\bibnamefont
  {Aguilar-Arevalo}} \emph {et~al.} (\bibinfo {collaboration} {PIENU}),\
  }\href@noop {} {\  (\bibinfo {year} {2019})},\ \Eprint
  {http://arxiv.org/abs/1904.03269} {arXiv:1904.03269 [hep-ex]} \BibitemShut
  {NoStop}%
\bibitem [{\citenamefont {Yamazaki}\ \emph {et~al.}(1984)\citenamefont
  {Yamazaki} \emph {et~al.}}]{Yamazaki:1984sj}%
  \BibitemOpen
  \bibfield  {author} {\bibinfo {author} {\bibfnamefont {T.}~\bibnamefont
  {Yamazaki}} \emph {et~al.},\ }\bibfield  {booktitle} {\emph {\bibinfo
  {booktitle} {{22nd International Conference on High Energy Physics. Vol. 1:
  Leipzig, Germany, July 19-25, 1984}}},\ }\href@noop {} {\ ,\ \bibinfo {pages}
  {I.262} (\bibinfo {year} {1984})},\ \bibinfo {note} {[Conf.
  Proc.C840719,262(1984)]}\BibitemShut {NoStop}%
\bibitem [{\citenamefont {Bergsma}\ \emph {et~al.}(1986)\citenamefont {Bergsma}
  \emph {et~al.}}]{Bergsma:1985is}%
  \BibitemOpen
  \bibfield  {author} {\bibinfo {author} {\bibfnamefont {F.}~\bibnamefont
  {Bergsma}} \emph {et~al.} (\bibinfo {collaboration} {CHARM}),\ }\href
  {\doibase 10.1016/0370-2693(86)91601-1} {\bibfield  {journal} {\bibinfo
  {journal} {Phys. Lett.}\ }\textbf {\bibinfo {volume} {166B}},\ \bibinfo
  {pages} {473} (\bibinfo {year} {1986})}\BibitemShut {NoStop}%
\bibitem [{\citenamefont {Abela}\ \emph {et~al.}(1981)\citenamefont {Abela},
  \citenamefont {Daum}, \citenamefont {Eaton}, \citenamefont {Frosch},
  \citenamefont {Jost}, \citenamefont {Kettle},\ and\ \citenamefont
  {Steiner}}]{Abela:1981nf}%
  \BibitemOpen
  \bibfield  {author} {\bibinfo {author} {\bibfnamefont {R.}~\bibnamefont
  {Abela}}, \bibinfo {author} {\bibfnamefont {M.}~\bibnamefont {Daum}},
  \bibinfo {author} {\bibfnamefont {G.~H.}\ \bibnamefont {Eaton}}, \bibinfo
  {author} {\bibfnamefont {R.}~\bibnamefont {Frosch}}, \bibinfo {author}
  {\bibfnamefont {B.}~\bibnamefont {Jost}}, \bibinfo {author} {\bibfnamefont
  {P.~R.}\ \bibnamefont {Kettle}}, \ and\ \bibinfo {author} {\bibfnamefont
  {E.}~\bibnamefont {Steiner}},\ }\href {\doibase 10.1016/0370-2693(81)90884-4}
  {\bibfield  {journal} {\bibinfo  {journal} {Phys. Lett.}\ }\textbf {\bibinfo
  {volume} {105B}},\ \bibinfo {pages} {263} (\bibinfo {year} {1981})},\
  \bibinfo {note} {[Erratum: Phys. Lett.106B,513(1981)]}\BibitemShut {NoStop}%
\bibitem [{\citenamefont {Bernardi}\ \emph {et~al.}(1986)\citenamefont
  {Bernardi} \emph {et~al.}}]{Bernardi:1985ny}%
  \BibitemOpen
  \bibfield  {author} {\bibinfo {author} {\bibfnamefont {G.}~\bibnamefont
  {Bernardi}} \emph {et~al.},\ }\href {\doibase 10.1016/0370-2693(86)91602-3}
  {\bibfield  {journal} {\bibinfo  {journal} {Phys. Lett. B}\ }\textbf
  {\bibinfo {volume} {166}},\ \bibinfo {pages} {479} (\bibinfo {year}
  {1986})}\BibitemShut {NoStop}%
\bibitem [{\citenamefont {Bernardi}\ \emph {et~al.}(1988)\citenamefont
  {Bernardi} \emph {et~al.}}]{Bernardi:1987ek}%
  \BibitemOpen
  \bibfield  {author} {\bibinfo {author} {\bibfnamefont {G.}~\bibnamefont
  {Bernardi}} \emph {et~al.},\ }\href {\doibase 10.1016/0370-2693(88)90563-1}
  {\bibfield  {journal} {\bibinfo  {journal} {Phys. Lett.}\ }\textbf {\bibinfo
  {volume} {B203}},\ \bibinfo {pages} {332} (\bibinfo {year}
  {1988})}\BibitemShut {NoStop}%
\bibitem [{\citenamefont {de~Gouv\^ea}\ and\ \citenamefont
  {Kobach}(2016)}]{deGouvea:2015euy}%
  \BibitemOpen
  \bibfield  {author} {\bibinfo {author} {\bibfnamefont {A.}~\bibnamefont
  {de~Gouv\^ea}}\ and\ \bibinfo {author} {\bibfnamefont {A.}~\bibnamefont
  {Kobach}},\ }\href {\doibase 10.1103/PhysRevD.93.033005} {\bibfield
  {journal} {\bibinfo  {journal} {Phys. Rev. D}\ }\textbf {\bibinfo {volume}
  {93}},\ \bibinfo {pages} {033005} (\bibinfo {year} {2016})},\ \Eprint
  {http://arxiv.org/abs/1511.00683} {arXiv:1511.00683 [hep-ph]} \BibitemShut
  {NoStop}%
\bibitem [{\citenamefont {Cortina~Gil}\ \emph {et~al.}(2020)\citenamefont
  {Cortina~Gil} \emph {et~al.}}]{NA62:2020mcv}%
  \BibitemOpen
  \bibfield  {author} {\bibinfo {author} {\bibfnamefont {E.}~\bibnamefont
  {Cortina~Gil}} \emph {et~al.} (\bibinfo {collaboration} {NA62}),\ }\href
  {\doibase 10.1016/j.physletb.2020.135599} {\bibfield  {journal} {\bibinfo
  {journal} {Phys. Lett. B}\ }\textbf {\bibinfo {volume} {807}},\ \bibinfo
  {pages} {135599} (\bibinfo {year} {2020})},\ \Eprint
  {http://arxiv.org/abs/2005.09575} {arXiv:2005.09575 [hep-ex]} \BibitemShut
  {NoStop}%
\bibitem [{\citenamefont {Bryman}\ and\ \citenamefont
  {Shrock}(2019{\natexlab{a}})}]{Bryman:2019ssi}%
  \BibitemOpen
  \bibfield  {author} {\bibinfo {author} {\bibfnamefont {D.~A.}\ \bibnamefont
  {Bryman}}\ and\ \bibinfo {author} {\bibfnamefont {R.}~\bibnamefont
  {Shrock}},\ }\href@noop {} {\  (\bibinfo {year} {2019}{\natexlab{a}})},\
  \Eprint {http://arxiv.org/abs/1904.06787} {arXiv:1904.06787 [hep-ph]}
  \BibitemShut {NoStop}%
\bibitem [{\citenamefont {Bryman}\ and\ \citenamefont
  {Shrock}(2019{\natexlab{b}})}]{Bryman:2019bjg}%
  \BibitemOpen
  \bibfield  {author} {\bibinfo {author} {\bibfnamefont {D.}~\bibnamefont
  {Bryman}}\ and\ \bibinfo {author} {\bibfnamefont {R.}~\bibnamefont
  {Shrock}},\ }\href {\doibase 10.1103/PhysRevD.100.073011} {\bibfield
  {journal} {\bibinfo  {journal} {Phys. Rev. D}\ }\textbf {\bibinfo {volume}
  {100}},\ \bibinfo {pages} {073011} (\bibinfo {year} {2019}{\natexlab{b}})},\
  \Eprint {http://arxiv.org/abs/1909.11198} {arXiv:1909.11198 [hep-ph]}
  \BibitemShut {NoStop}%
\bibitem [{\citenamefont {Dolgov}\ \emph
  {et~al.}(2000{\natexlab{a}})\citenamefont {Dolgov}, \citenamefont {Hansen},
  \citenamefont {Raffelt},\ and\ \citenamefont {Semikoz}}]{Dolgov:2000jw}%
  \BibitemOpen
  \bibfield  {author} {\bibinfo {author} {\bibfnamefont {A.~D.}\ \bibnamefont
  {Dolgov}}, \bibinfo {author} {\bibfnamefont {S.~H.}\ \bibnamefont {Hansen}},
  \bibinfo {author} {\bibfnamefont {G.}~\bibnamefont {Raffelt}}, \ and\
  \bibinfo {author} {\bibfnamefont {D.~V.}\ \bibnamefont {Semikoz}},\ }\href
  {\doibase 10.1016/S0550-3213(00)00566-6} {\bibfield  {journal} {\bibinfo
  {journal} {Nucl. Phys.}\ }\textbf {\bibinfo {volume} {B590}},\ \bibinfo
  {pages} {562} (\bibinfo {year} {2000}{\natexlab{a}})},\ \Eprint
  {http://arxiv.org/abs/hep-ph/0008138} {arXiv:hep-ph/0008138 [hep-ph]}
  \BibitemShut {NoStop}%
\bibitem [{\citenamefont {Ruchayskiy}\ and\ \citenamefont
  {Ivashko}(2012)}]{Ruchayskiy:2012si}%
  \BibitemOpen
  \bibfield  {author} {\bibinfo {author} {\bibfnamefont {O.}~\bibnamefont
  {Ruchayskiy}}\ and\ \bibinfo {author} {\bibfnamefont {A.}~\bibnamefont
  {Ivashko}},\ }\href {\doibase 10.1088/1475-7516/2012/10/014} {\bibfield
  {journal} {\bibinfo  {journal} {JCAP}\ }\textbf {\bibinfo {volume} {1210}},\
  \bibinfo {pages} {014} (\bibinfo {year} {2012})},\ \Eprint
  {http://arxiv.org/abs/1202.2841} {arXiv:1202.2841 [hep-ph]} \BibitemShut
  {NoStop}%
\bibitem [{\citenamefont {Boyarsky}\ \emph {et~al.}(2020)\citenamefont
  {Boyarsky}, \citenamefont {Ovchynnikov}, \citenamefont {Ruchayskiy},\ and\
  \citenamefont {Syvolap}}]{Boyarsky:2020dzc}%
  \BibitemOpen
  \bibfield  {author} {\bibinfo {author} {\bibfnamefont {A.}~\bibnamefont
  {Boyarsky}}, \bibinfo {author} {\bibfnamefont {M.}~\bibnamefont
  {Ovchynnikov}}, \bibinfo {author} {\bibfnamefont {O.}~\bibnamefont
  {Ruchayskiy}}, \ and\ \bibinfo {author} {\bibfnamefont {V.}~\bibnamefont
  {Syvolap}},\ }\href@noop {} {\  (\bibinfo {year} {2020})},\ \Eprint
  {http://arxiv.org/abs/2008.00749} {arXiv:2008.00749 [hep-ph]} \BibitemShut
  {NoStop}%
\bibitem [{\citenamefont {Bondarenko}\ \emph {et~al.}(2021)\citenamefont
  {Bondarenko}, \citenamefont {Boyarsky}, \citenamefont {Klaric}, \citenamefont
  {Mikulenko}, \citenamefont {Ruchayskiy}, \citenamefont {Syvolap},\ and\
  \citenamefont {Timiryasov}}]{Bondarenko:2021cpc}%
  \BibitemOpen
  \bibfield  {author} {\bibinfo {author} {\bibfnamefont {K.}~\bibnamefont
  {Bondarenko}}, \bibinfo {author} {\bibfnamefont {A.}~\bibnamefont
  {Boyarsky}}, \bibinfo {author} {\bibfnamefont {J.}~\bibnamefont {Klaric}},
  \bibinfo {author} {\bibfnamefont {O.}~\bibnamefont {Mikulenko}}, \bibinfo
  {author} {\bibfnamefont {O.}~\bibnamefont {Ruchayskiy}}, \bibinfo {author}
  {\bibfnamefont {V.}~\bibnamefont {Syvolap}}, \ and\ \bibinfo {author}
  {\bibfnamefont {I.}~\bibnamefont {Timiryasov}},\ }\href@noop {} {\  (\bibinfo
  {year} {2021})},\ \Eprint {http://arxiv.org/abs/2101.09255} {arXiv:2101.09255
  [hep-ph]} \BibitemShut {NoStop}%
\bibitem [{\citenamefont {Fuller}\ \emph {et~al.}(2011)\citenamefont {Fuller},
  \citenamefont {Kishimoto},\ and\ \citenamefont {Kusenko}}]{Fuller:2011qy}%
  \BibitemOpen
  \bibfield  {author} {\bibinfo {author} {\bibfnamefont {G.~M.}\ \bibnamefont
  {Fuller}}, \bibinfo {author} {\bibfnamefont {C.~T.}\ \bibnamefont
  {Kishimoto}}, \ and\ \bibinfo {author} {\bibfnamefont {A.}~\bibnamefont
  {Kusenko}},\ }\href@noop {} {\  (\bibinfo {year} {2011})},\ \Eprint
  {http://arxiv.org/abs/1110.6479} {arXiv:1110.6479 [astro-ph.CO]} \BibitemShut
  {NoStop}%
\bibitem [{\citenamefont {Atre}\ \emph {et~al.}(2009)\citenamefont {Atre},
  \citenamefont {Han}, \citenamefont {Pascoli},\ and\ \citenamefont
  {Zhang}}]{Atre:2009rg}%
  \BibitemOpen
  \bibfield  {author} {\bibinfo {author} {\bibfnamefont {A.}~\bibnamefont
  {Atre}}, \bibinfo {author} {\bibfnamefont {T.}~\bibnamefont {Han}}, \bibinfo
  {author} {\bibfnamefont {S.}~\bibnamefont {Pascoli}}, \ and\ \bibinfo
  {author} {\bibfnamefont {B.}~\bibnamefont {Zhang}},\ }\href {\doibase
  10.1088/1126-6708/2009/05/030} {\bibfield  {journal} {\bibinfo  {journal}
  {JHEP}\ }\textbf {\bibinfo {volume} {05}},\ \bibinfo {pages} {030} (\bibinfo
  {year} {2009})},\ \Eprint {http://arxiv.org/abs/0901.3589} {arXiv:0901.3589
  [hep-ph]} \BibitemShut {NoStop}%
\bibitem [{\citenamefont {Benes}\ \emph {et~al.}(2005)\citenamefont {Benes},
  \citenamefont {Faessler}, \citenamefont {Simkovic},\ and\ \citenamefont
  {Kovalenko}}]{Benes:2005hn}%
  \BibitemOpen
  \bibfield  {author} {\bibinfo {author} {\bibfnamefont {P.}~\bibnamefont
  {Benes}}, \bibinfo {author} {\bibfnamefont {A.}~\bibnamefont {Faessler}},
  \bibinfo {author} {\bibfnamefont {F.}~\bibnamefont {Simkovic}}, \ and\
  \bibinfo {author} {\bibfnamefont {S.}~\bibnamefont {Kovalenko}},\ }\href
  {\doibase 10.1103/PhysRevD.71.077901} {\bibfield  {journal} {\bibinfo
  {journal} {Phys. Rev.}\ }\textbf {\bibinfo {volume} {D71}},\ \bibinfo {pages}
  {077901} (\bibinfo {year} {2005})},\ \Eprint
  {http://arxiv.org/abs/hep-ph/0501295} {arXiv:hep-ph/0501295 [hep-ph]}
  \BibitemShut {NoStop}%
\bibitem [{\citenamefont {Kusenko}\ \emph {et~al.}(2005)\citenamefont
  {Kusenko}, \citenamefont {Pascoli},\ and\ \citenamefont
  {Semikoz}}]{Kusenko:2004qc}%
  \BibitemOpen
  \bibfield  {author} {\bibinfo {author} {\bibfnamefont {A.}~\bibnamefont
  {Kusenko}}, \bibinfo {author} {\bibfnamefont {S.}~\bibnamefont {Pascoli}}, \
  and\ \bibinfo {author} {\bibfnamefont {D.}~\bibnamefont {Semikoz}},\ }\href
  {\doibase 10.1088/1126-6708/2005/11/028} {\bibfield  {journal} {\bibinfo
  {journal} {JHEP}\ }\textbf {\bibinfo {volume} {11}},\ \bibinfo {pages} {028}
  (\bibinfo {year} {2005})},\ \Eprint {http://arxiv.org/abs/hep-ph/0405198}
  {arXiv:hep-ph/0405198 [hep-ph]} \BibitemShut {NoStop}%
\bibitem [{\citenamefont {Ballett}\ \emph {et~al.}(2020)\citenamefont
  {Ballett}, \citenamefont {Boschi},\ and\ \citenamefont
  {Pascoli}}]{Ballett:2019bgd}%
  \BibitemOpen
  \bibfield  {author} {\bibinfo {author} {\bibfnamefont {P.}~\bibnamefont
  {Ballett}}, \bibinfo {author} {\bibfnamefont {T.}~\bibnamefont {Boschi}}, \
  and\ \bibinfo {author} {\bibfnamefont {S.}~\bibnamefont {Pascoli}},\ }\href
  {\doibase 10.1007/JHEP03(2020)111} {\bibfield  {journal} {\bibinfo  {journal}
  {JHEP}\ }\textbf {\bibinfo {volume} {03}},\ \bibinfo {pages} {111} (\bibinfo
  {year} {2020})},\ \Eprint {http://arxiv.org/abs/1905.00284} {arXiv:1905.00284
  [hep-ph]} \BibitemShut {NoStop}%
\bibitem [{\citenamefont {Dolgov}\ \emph
  {et~al.}(2000{\natexlab{b}})\citenamefont {Dolgov}, \citenamefont {Hansen},
  \citenamefont {Raffelt},\ and\ \citenamefont {Semikoz}}]{Dolgov:2000pj}%
  \BibitemOpen
  \bibfield  {author} {\bibinfo {author} {\bibfnamefont {A.~D.}\ \bibnamefont
  {Dolgov}}, \bibinfo {author} {\bibfnamefont {S.~H.}\ \bibnamefont {Hansen}},
  \bibinfo {author} {\bibfnamefont {G.}~\bibnamefont {Raffelt}}, \ and\
  \bibinfo {author} {\bibfnamefont {D.~V.}\ \bibnamefont {Semikoz}},\ }\href
  {\doibase 10.1016/S0550-3213(00)00203-0} {\bibfield  {journal} {\bibinfo
  {journal} {Nucl. Phys.}\ }\textbf {\bibinfo {volume} {B580}},\ \bibinfo
  {pages} {331} (\bibinfo {year} {2000}{\natexlab{b}})},\ \Eprint
  {http://arxiv.org/abs/hep-ph/0002223} {arXiv:hep-ph/0002223 [hep-ph]}
  \BibitemShut {NoStop}%
\bibitem [{\citenamefont {Mastrototaro}\ \emph {et~al.}(2020)\citenamefont
  {Mastrototaro}, \citenamefont {Mirizzi}, \citenamefont {Serpico},\ and\
  \citenamefont {Esmaili}}]{Mastrototaro:2019vug}%
  \BibitemOpen
  \bibfield  {author} {\bibinfo {author} {\bibfnamefont {L.}~\bibnamefont
  {Mastrototaro}}, \bibinfo {author} {\bibfnamefont {A.}~\bibnamefont
  {Mirizzi}}, \bibinfo {author} {\bibfnamefont {P.~D.}\ \bibnamefont
  {Serpico}}, \ and\ \bibinfo {author} {\bibfnamefont {A.}~\bibnamefont
  {Esmaili}},\ }\href {\doibase 10.1088/1475-7516/2020/01/010} {\bibfield
  {journal} {\bibinfo  {journal} {JCAP}\ }\textbf {\bibinfo {volume} {01}},\
  \bibinfo {pages} {010} (\bibinfo {year} {2020})},\ \Eprint
  {http://arxiv.org/abs/1910.10249} {arXiv:1910.10249 [hep-ph]} \BibitemShut
  {NoStop}%
\bibitem [{\citenamefont {Syvolap}\ \emph {et~al.}(2019)\citenamefont
  {Syvolap}, \citenamefont {Ruchayskiy},\ and\ \citenamefont
  {Boyarsky}}]{Syvolap:2019dat}%
  \BibitemOpen
  \bibfield  {author} {\bibinfo {author} {\bibfnamefont {V.}~\bibnamefont
  {Syvolap}}, \bibinfo {author} {\bibfnamefont {O.}~\bibnamefont {Ruchayskiy}},
  \ and\ \bibinfo {author} {\bibfnamefont {A.}~\bibnamefont {Boyarsky}},\
  }\href@noop {} {\  (\bibinfo {year} {2019})},\ \Eprint
  {http://arxiv.org/abs/1909.06320} {arXiv:1909.06320 [hep-ph]} \BibitemShut
  {NoStop}%
\bibitem [{\citenamefont {Suliga}\ \emph {et~al.}(2019)\citenamefont {Suliga},
  \citenamefont {Tamborra},\ and\ \citenamefont {Wu}}]{Suliga:2019bsq}%
  \BibitemOpen
  \bibfield  {author} {\bibinfo {author} {\bibfnamefont {A.~M.}\ \bibnamefont
  {Suliga}}, \bibinfo {author} {\bibfnamefont {I.}~\bibnamefont {Tamborra}}, \
  and\ \bibinfo {author} {\bibfnamefont {M.-R.}\ \bibnamefont {Wu}},\ }\href
  {\doibase 10.1088/1475-7516/2019/12/019} {\bibfield  {journal} {\bibinfo
  {journal} {JCAP}\ }\textbf {\bibinfo {volume} {12}},\ \bibinfo {pages} {019}
  (\bibinfo {year} {2019})},\ \Eprint {http://arxiv.org/abs/1908.11382}
  {arXiv:1908.11382 [astro-ph.HE]} \BibitemShut {NoStop}%
\bibitem [{\citenamefont {Suliga}\ \emph {et~al.}(2020)\citenamefont {Suliga},
  \citenamefont {Tamborra},\ and\ \citenamefont {Wu}}]{Suliga:2020vpz}%
  \BibitemOpen
  \bibfield  {author} {\bibinfo {author} {\bibfnamefont {A.~M.}\ \bibnamefont
  {Suliga}}, \bibinfo {author} {\bibfnamefont {I.}~\bibnamefont {Tamborra}}, \
  and\ \bibinfo {author} {\bibfnamefont {M.-R.}\ \bibnamefont {Wu}},\ }\href
  {\doibase 10.1088/1475-7516/2020/08/018} {\bibfield  {journal} {\bibinfo
  {journal} {JCAP}\ }\textbf {\bibinfo {volume} {08}},\ \bibinfo {pages} {018}
  (\bibinfo {year} {2020})},\ \Eprint {http://arxiv.org/abs/2004.11389}
  {arXiv:2004.11389 [astro-ph.HE]} \BibitemShut {NoStop}%
\bibitem [{\citenamefont {Izotov}\ \emph {et~al.}(2014)\citenamefont {Izotov},
  \citenamefont {Thuan},\ and\ \citenamefont {Guseva}}]{Izotov:2014fga}%
  \BibitemOpen
  \bibfield  {author} {\bibinfo {author} {\bibfnamefont {Y.~I.}\ \bibnamefont
  {Izotov}}, \bibinfo {author} {\bibfnamefont {T.~X.}\ \bibnamefont {Thuan}}, \
  and\ \bibinfo {author} {\bibfnamefont {N.~G.}\ \bibnamefont {Guseva}},\
  }\href {\doibase 10.1093/mnras/stu1771} {\bibfield  {journal} {\bibinfo
  {journal} {Mon. Not. Roy. Astron. Soc.}\ }\textbf {\bibinfo {volume} {445}},\
  \bibinfo {pages} {778} (\bibinfo {year} {2014})},\ \Eprint
  {http://arxiv.org/abs/1408.6953} {arXiv:1408.6953 [astro-ph.CO]} \BibitemShut
  {NoStop}%
\bibitem [{\citenamefont {Tanabashi}\ \emph {et~al.}(2018)\citenamefont
  {Tanabashi} \emph {et~al.}}]{Tanabashi:2018oca}%
  \BibitemOpen
  \bibfield  {author} {\bibinfo {author} {\bibfnamefont {M.}~\bibnamefont
  {Tanabashi}} \emph {et~al.} (\bibinfo {collaboration} {Particle Data
  Group}),\ }\href {\doibase 10.1103/PhysRevD.98.030001} {\bibfield  {journal}
  {\bibinfo  {journal} {Phys. Rev.}\ }\textbf {\bibinfo {volume} {D98}},\
  \bibinfo {pages} {030001} (\bibinfo {year} {2018})}\BibitemShut {NoStop}%
\bibitem [{\citenamefont {{Zavarygin}}\ \emph {et~al.}(2018)\citenamefont
  {{Zavarygin}}, \citenamefont {{Webb}}, \citenamefont {{Dumont}},\ and\
  \citenamefont {{Riemer-S{\o}rensen}}}]{Zavarygin:2017}%
  \BibitemOpen
  \bibfield  {author} {\bibinfo {author} {\bibfnamefont {E.~O.}\ \bibnamefont
  {{Zavarygin}}}, \bibinfo {author} {\bibfnamefont {J.~K.}\ \bibnamefont
  {{Webb}}}, \bibinfo {author} {\bibfnamefont {V.}~\bibnamefont {{Dumont}}}, \
  and\ \bibinfo {author} {\bibfnamefont {S.}~\bibnamefont
  {{Riemer-S{\o}rensen}}},\ }\href {\doibase 10.1093/mnras/sty1003} {\bibfield
  {journal} {\bibinfo  {journal} {Mon. Not. Roy. Astron. Soc.}\ }\textbf
  {\bibinfo {volume} {477}},\ \bibinfo {pages} {5536} (\bibinfo {year}
  {2018})},\ \Eprint {http://arxiv.org/abs/1706.09512} {arXiv:1706.09512
  [astro-ph.GA]} \BibitemShut {NoStop}%
\bibitem [{\citenamefont {Cyburt}\ \emph {et~al.}(2016)\citenamefont {Cyburt},
  \citenamefont {Fields}, \citenamefont {Olive},\ and\ \citenamefont
  {Yeh}}]{Cyburt:2015mya}%
  \BibitemOpen
  \bibfield  {author} {\bibinfo {author} {\bibfnamefont {R.~H.}\ \bibnamefont
  {Cyburt}}, \bibinfo {author} {\bibfnamefont {B.~D.}\ \bibnamefont {Fields}},
  \bibinfo {author} {\bibfnamefont {K.~A.}\ \bibnamefont {Olive}}, \ and\
  \bibinfo {author} {\bibfnamefont {T.-H.}\ \bibnamefont {Yeh}},\ }\href
  {\doibase 10.1103/RevModPhys.88.015004} {\bibfield  {journal} {\bibinfo
  {journal} {Rev. Mod. Phys.}\ }\textbf {\bibinfo {volume} {88}},\ \bibinfo
  {pages} {015004} (\bibinfo {year} {2016})},\ \Eprint
  {http://arxiv.org/abs/1505.01076} {arXiv:1505.01076 [astro-ph.CO]}
  \BibitemShut {NoStop}%
\bibitem [{\citenamefont {Lesgourgues}\ and\ \citenamefont
  {Pastor}(2006)}]{Lesgourgues:2006nd}%
  \BibitemOpen
  \bibfield  {author} {\bibinfo {author} {\bibfnamefont {J.}~\bibnamefont
  {Lesgourgues}}\ and\ \bibinfo {author} {\bibfnamefont {S.}~\bibnamefont
  {Pastor}},\ }\href {\doibase 10.1016/j.physrep.2006.04.001} {\bibfield
  {journal} {\bibinfo  {journal} {Phys. Rept.}\ }\textbf {\bibinfo {volume}
  {429}},\ \bibinfo {pages} {307} (\bibinfo {year} {2006})},\ \Eprint
  {http://arxiv.org/abs/astro-ph/0603494} {arXiv:astro-ph/0603494 [astro-ph]}
  \BibitemShut {NoStop}%
\bibitem [{\citenamefont {Font-Ribera}\ \emph {et~al.}(2014)\citenamefont
  {Font-Ribera}, \citenamefont {McDonald}, \citenamefont {Mostek},
  \citenamefont {Reid}, \citenamefont {Seo},\ and\ \citenamefont
  {Slosar}}]{Font-Ribera:2013rwa}%
  \BibitemOpen
  \bibfield  {author} {\bibinfo {author} {\bibfnamefont {A.}~\bibnamefont
  {Font-Ribera}}, \bibinfo {author} {\bibfnamefont {P.}~\bibnamefont
  {McDonald}}, \bibinfo {author} {\bibfnamefont {N.}~\bibnamefont {Mostek}},
  \bibinfo {author} {\bibfnamefont {B.~A.}\ \bibnamefont {Reid}}, \bibinfo
  {author} {\bibfnamefont {H.-J.}\ \bibnamefont {Seo}}, \ and\ \bibinfo
  {author} {\bibfnamefont {A.}~\bibnamefont {Slosar}},\ }\href {\doibase
  10.1088/1475-7516/2014/05/023} {\bibfield  {journal} {\bibinfo  {journal}
  {JCAP}\ }\textbf {\bibinfo {volume} {1405}},\ \bibinfo {pages} {023}
  (\bibinfo {year} {2014})},\ \Eprint {http://arxiv.org/abs/1308.4164}
  {arXiv:1308.4164 [astro-ph.CO]} \BibitemShut {NoStop}%
\bibitem [{\citenamefont {Bolliet}\ \emph {et~al.}(2019)\citenamefont
  {Bolliet}, \citenamefont {Brinckmann}, \citenamefont {Chluba},\ and\
  \citenamefont {Lesgourgues}}]{Bolliet:2019zuz}%
  \BibitemOpen
  \bibfield  {author} {\bibinfo {author} {\bibfnamefont {B.}~\bibnamefont
  {Bolliet}}, \bibinfo {author} {\bibfnamefont {T.}~\bibnamefont {Brinckmann}},
  \bibinfo {author} {\bibfnamefont {J.}~\bibnamefont {Chluba}}, \ and\ \bibinfo
  {author} {\bibfnamefont {J.}~\bibnamefont {Lesgourgues}},\ }\href@noop {} {\
  (\bibinfo {year} {2019})},\ \Eprint {http://arxiv.org/abs/1906.10359}
  {arXiv:1906.10359 [astro-ph.CO]} \BibitemShut {NoStop}%
\bibitem [{\citenamefont {Gelmini}\ \emph
  {et~al.}(2020{\natexlab{c}})\citenamefont {Gelmini}, \citenamefont
  {Kawasaki}, \citenamefont {Kusenko}, \citenamefont {Murai},\ and\
  \citenamefont {Takhistov}}]{Gelmini:2020ekg}%
  \BibitemOpen
  \bibfield  {author} {\bibinfo {author} {\bibfnamefont {G.~B.}\ \bibnamefont
  {Gelmini}}, \bibinfo {author} {\bibfnamefont {M.}~\bibnamefont {Kawasaki}},
  \bibinfo {author} {\bibfnamefont {A.}~\bibnamefont {Kusenko}}, \bibinfo
  {author} {\bibfnamefont {K.}~\bibnamefont {Murai}}, \ and\ \bibinfo {author}
  {\bibfnamefont {V.}~\bibnamefont {Takhistov}},\ }\href@noop {} {\  (\bibinfo
  {year} {2020}{\natexlab{c}})},\ \Eprint {http://arxiv.org/abs/2005.06721}
  {arXiv:2005.06721 [hep-ph]} \BibitemShut {NoStop}%
\bibitem [{\citenamefont {Boyarsky}\ \emph {et~al.}(2021)\citenamefont
  {Boyarsky}, \citenamefont {Ovchynnikov}, \citenamefont {Sabti},\ and\
  \citenamefont {Syvolap}}]{Boyarsky:2021yoh}%
  \BibitemOpen
  \bibfield  {author} {\bibinfo {author} {\bibfnamefont {A.}~\bibnamefont
  {Boyarsky}}, \bibinfo {author} {\bibfnamefont {M.}~\bibnamefont
  {Ovchynnikov}}, \bibinfo {author} {\bibfnamefont {N.}~\bibnamefont {Sabti}},
  \ and\ \bibinfo {author} {\bibfnamefont {V.}~\bibnamefont {Syvolap}},\
  }\href@noop {} {\  (\bibinfo {year} {2021})},\ \Eprint
  {http://arxiv.org/abs/2103.09831} {arXiv:2103.09831 [hep-ph]} \BibitemShut
  {NoStop}%
\bibitem [{\citenamefont {Mastrototaro}\ \emph {et~al.}(2021)\citenamefont
  {Mastrototaro}, \citenamefont {Serpico}, \citenamefont {Mirizzi},\ and\
  \citenamefont {Saviano}}]{Mastrototaro:2021wzl}%
  \BibitemOpen
  \bibfield  {author} {\bibinfo {author} {\bibfnamefont {L.}~\bibnamefont
  {Mastrototaro}}, \bibinfo {author} {\bibfnamefont {P.~D.}\ \bibnamefont
  {Serpico}}, \bibinfo {author} {\bibfnamefont {A.}~\bibnamefont {Mirizzi}}, \
  and\ \bibinfo {author} {\bibfnamefont {N.}~\bibnamefont {Saviano}},\
  }\href@noop {} {\  (\bibinfo {year} {2021})},\ \Eprint
  {http://arxiv.org/abs/2104.11752} {arXiv:2104.11752 [hep-ph]} \BibitemShut
  {NoStop}%
\bibitem [{\citenamefont {Abazajian}\ \emph {et~al.}(2016)\citenamefont
  {Abazajian} \emph {et~al.}}]{Abazajian:2016yjj}%
  \BibitemOpen
  \bibfield  {author} {\bibinfo {author} {\bibfnamefont {K.~N.}\ \bibnamefont
  {Abazajian}} \emph {et~al.} (\bibinfo {collaboration} {CMB-S4}),\ }\href@noop
  {} {\  (\bibinfo {year} {2016})},\ \Eprint {http://arxiv.org/abs/1610.02743}
  {arXiv:1610.02743 [astro-ph.CO]} \BibitemShut {NoStop}%
\end{thebibliography}%
\end{document}